\documentclass[apj,twocolappendix,numberedappendix,appendixfloats]{emulateapj}

\usepackage{natbib}
\usepackage{amsmath}
\usepackage{xcolor}
\usepackage[english]{babel}
\usepackage{graphicx}
\usepackage{xspace}

\newcommand{\hMpc}{{\ifmmode{h^{-1}{\rm Mpc}}\else{$h^{-1}$Mpc}\fi}}
\newcommand{\Mpc}{{\ifmmode{{\rm Mpc}}\else{Mpc}\fi}}
\newcommand{\hkpc}{{\ifmmode{h^{-1}{\rm kpc}}\else{$h^{-1}$kpc}\fi}}
\newcommand{\kpc}{{\ifmmode{ {\rm kpc} }\else{{\rm kpc}}\fi}}
\newcommand{\kms}{{\ifmmode{ {\rm km\,s^{-1}} }\else{ ${\rm km\,s^{-1}}$ }\fi}}
\newcommand{\hMsun}{{\ifmmode{h^{-1}{\rm {M_{\odot}}}}\else{$h^{-1}{\rm{M_{\odot}}}$}\fi}}
\newcommand{\Msun}{{\ifmmode{{\rm M}_{\odot}}\else{${\rm M}_{\odot}$}\fi}}
\newcommand{\Mhalo}{{\ifmmode{M_{\rm halo}}\else{$M_{\rm halo}$}\fi}}
\newcommand{\Rvir}{{\ifmmode{R_{\rm vir}}\else{$R_{\rm vir}$}\fi}}
\newcommand{\Mvir}{{\ifmmode{M_{\rm vir}}\else{$M_{\rm vir}$}\fi}}
\newcommand{\Mstar}{{\ifmmode{M_{\rm star}}\else{$M_{\rm star}$}\fi}}
\newcommand{\Vrot}{{\ifmmode{V_{\rm rot}}\else{$V_{\rm rot}$}\fi}}
\newcommand{\ltsima}{$\; \buildrel < \over \sim \;$}
\newcommand{\gtsima}{$\; \buildrel > \over \sim \;$}
\newcommand{\lsim}{\lower.5ex\hbox{\ltsima}}
\newcommand{\gsim}{\lower.5ex\hbox{\gtsima}}

\def\lesssim{\mathrel{\hbox{\rlap{\hbox{\lower4pt\hbox{$\sim$}}}\hbox{$<$}}}}
\def\gtrsim{\mathrel{\hbox{\rlap{\hbox{\lower4pt\hbox{$\sim$}}}\hbox{$>$}}}}

\newcommand{\beq}{\begin{equation}}
\newcommand{\eeq}{\end{equation}}
\def\beqa{\begin{eqnarray}}
\def\eeqa{\end{eqnarray}}
\def\LCDM{\ensuremath{\Lambda}CDM}

\def\head{ \vbox to 0pt{\vss \hbox to 0pt{\hskip 440pt\rm
      LA-UR-10-07069\hss} \vskip 25pt}}

\def \kms {\ifmmode  \,\rm km\,s^{-1} \else $\,\rm km\,s^{-1}  $ \fi }
\def \kpc {\ifmmode  {\rm kpc}  \else ${\rm  kpc}$ \fi  }  
\def \hkpc {\ifmmode  {h^{-1}\rm kpc}  \else ${h^{-1}\rm kpc}$ \fi  }  
\def \hMpc {\ifmmode  {h^{-1}\rm Mpc}  \else ${h^{-1}\rm Mpc}$ \fi  }  
\def \Mpch {\ifmmode  {h^{-1}\rm Mpc}  \else ${h^{-1}\rm Mpc}$ \fi  }  
\def \Msun {\ifmmode {\rm M}_{\odot} \else ${\rm M}_{\odot}$ \fi} 
\def \hMsun {\ifmmode h^{-1}\,\rm M_{\odot} \else $h^{-1}\,\rm M_{\odot}$ \fi}

\def \LCDM {\ifmmode \Lambda{\rm CDM} \else $\Lambda{\rm CDM}$ \fi}
\def \sig8 {\ifmmode \sigma_8 \else $\sigma_8$ \fi} 
\def \OmegaM {\ifmmode \Omega_{\rm m} \else $\Omega_{\rm m}$ \fi} 
\def \Omegab {\ifmmode \Omega_{\rm b} \else $\Omega_{\rm b}$ \fi} 
\def \OmegaL {\ifmmode \Omega_{\rm \Lambda} \else $\Omega_{\rm \Lambda}$\fi} 
\def \Deltavir {\ifmmode \Delta_{\rm vir} \else $\Delta_{\rm vir}$ \fi}
\def \rhocrit {\ifmmode \rho_{\rm crit} \else $\rho_{\rm crit}$ \fi}
\def \rhou {\ifmmode \rho_{\rm u} \else $\rho_{\rm u}$ \fi}
\def \zc {\ifmmode z_{\rm c} \else $z_{\rm c}$ \fi}

\slugcomment{accepted by ApJ}
\shorttitle{The formation of the MW bulge}
\shortauthors{Buck, Ness, Obreja, et al.}

\begin{document}

\title{Stars behind bars II: A cosmological formation scenario for the Milky Way's central stellar structure}

\author{Tobias Buck\altaffilmark{1}\altaffilmark{$\star$}\altaffilmark{$\dagger$}, Melissa Ness\altaffilmark{2,3}, Aura Obreja\altaffilmark{4}, Andrea V. Macci\`o\altaffilmark{5,1}, Aaron A. Dutton\altaffilmark{5}}
%\altaffilmark{$\ddagger$}
\altaffiltext{$\star$}{buck@mpia.de}
%\altaffiltext{$\ddagger$}{maccio@nyu.edu}
\altaffiltext{1}{Max-Planck Institut f\"ur Astronomie, K\"onigstuhl 17, D-69117 Heidelberg, Germany}
\altaffiltext{2}{Department of Astronomy, Columbia University, Pupin Physics Laboratories, New York, NY 10027, USA}\
\altaffiltext{3}{Center for Computational Astrophysics, Flatiron Institute, 162 Fifth Avenue, New York, NY 10010, USA}
\altaffiltext{4}{Universit\"ats-Sternwarte M\"unchen, Scheinerstraße 1, D-81679 M\"unchen, Germany}
\altaffiltext{5}{New York University Abu Dhabi, PO Box 129188, Abu Dhabi, UAE}

\altaffiltext{$\dagger$}{Member of the International Max Planck Research School for Astronomy and Cosmic Physics at the University of Heidelberg, IMPRS-HD, Germany.}

\begin{abstract}
The stellar populations in the inner kiloparsecs of the Milky Way (MW) show complex kinematical and chemical structures. The origin and evolution of these structures is still under debate. Here we study the central region of a \textit{fully cosmological hydrodynamical} simulation of a disk galaxy that reproduces key properties of the inner kiloparsecs of the MW: it has a boxy morphology and shows an overall rotation and dispersion profile in agreement with observations. 
We use a clustering algorithm on stellar kinematics to identify a number of discrete kinematic components: a high- and low-spin disk, a stellar halo and two bulge components; one fast rotating and one \textbf{slow-rotating}. We focus on the two bulge components and show that the slow rotating one is spherically symmetric while the fast rotating component shows a boxy/peanut morphology.  Although the two bulge components are kinematically discrete populations at present-day, they are \textit{both} mostly formed over similar time scales, from disk material. We find that stellar particles  with lower initial birth angular momentum (most likely thick disc stars) end up in the \textbf{slow-rotating} low-spin bulge, while stars with higher birth angular momentum (most likely thin disc stars) are found in the high-spin bulge. This has the important consequence that a bulge population with a spheroidal morphology does not necessarily indicate a merger origin. In fact, we do find that only $\sim2.3$\% of the stars in the bulge components are ex-situ stars brought in by accreted dwarf galaxies early on. We identify these ex-situ stars as the oldest and most metal-poor stars on highly radial orbits with large vertical excursions from the disk. 
\end{abstract}

%%%%%%%%%%%%%%%%%%%%%%%%%%%%%%%%%%%%%%%%%%%%%%%%%%%
\section{Introduction} \label{sec:introduction}
%%%%%%%%%%%%%%%%%%%%%%%%%%%%%%%%%%%%%%%%%%%%%%%%%%%

%%%%%%%%%%%%%%%%%% FIGURE 1 %%%%%%%%%%%%%%%%%%%%%%%%%%%
\begin{figure*}
\begin{center}
\includegraphics[trim={0 1.75cm 0 1.75cm},clip,width=.75\textwidth]{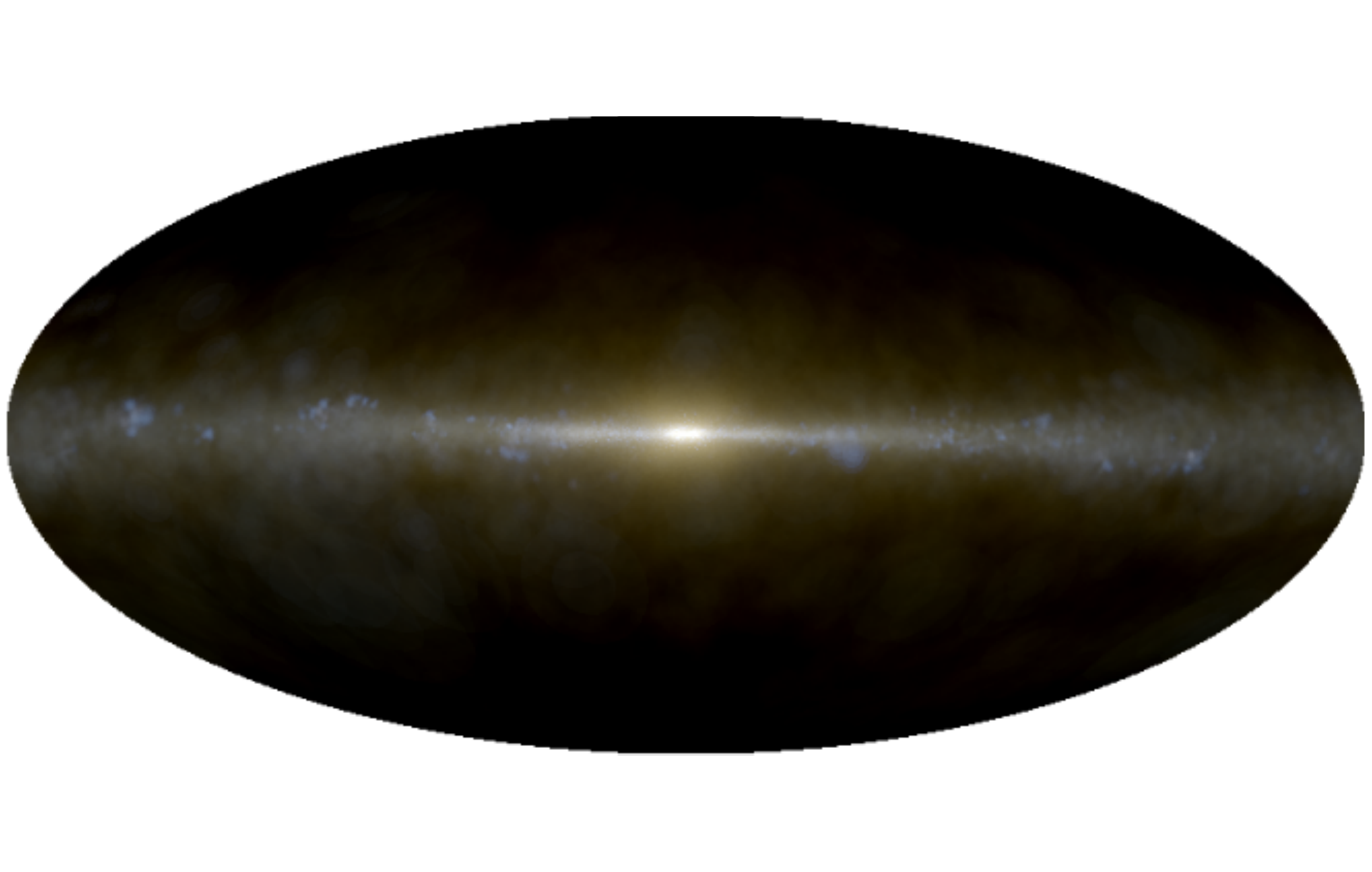}
\end{center}
%\vspace{-.35cm}
\caption{RGB all sky image of the simulated galaxy in an Aitoff projection. The camera position is chosen to be 8 kpc from the Galactic center imitating the earth's position in the Milky Way.
We render the stellar particles using their luminosity in the wavelength bands $i,v,u$ in order to create the r, g and b channels.
}
\label{fig:all_sky}
\end{figure*}
%%%%%%%%%%%%%%%%%%%%%%%%%%%%%%%%%%%%%%%%%%%%%%%%%%%%

The structure and dynamics of stars in the inner region of disk galaxies carries important information about the formation and subsequent evolution processes shaping them. While the inner parts of disk galaxies can have a boxy or spheroidal morphology, several stellar constituents can co-exist, including halo, disk, bulge and cluster populations. Some galaxies, like M31, show evidence for multiple bulge populations, of both a spherical and boxy component \citep{Beaton2007,Blana2017}.  Typically, boxy bulges are attributed to internal evolution processes, formed from the disk \citep[e.g.][]{Combes1981,Raha1991,Bureau1999}. These processes are driven by the influence of non-axisymmetric components, such as the bar \citep[e.g.][]{Herpich2017,Fragkoudi2018}.  Conversely, large spheroidal bulge components are typically attributed to turbulent, external formation processes, such as mergers.  

About 50 percent of all nearby galaxies, including the Milky Way (hereafter, MW) have a boxy or peanut shaped bulge \citep[e.g.][]{Lutticke2004,Okuda1977,Blitz1991,Gonzalez2017}. Over the last two decades, a large number of studies have explored the structure of the MW's bulge, revealing the presence of rich substructures within the inner region \citep[e.g.][]{Zoccali2008,Lecureur2007,Babusiaux2010,Gonzalez2013,Bensby2013,Zasowski2016}. The BRAVA \citep{Howard2008}, ARGOS \citep{Freeman2013} and APOGEE survey data \citep{Apogee} show that the stars in the bulge of the MW are cylindrically rotating \citep{Howard2009,Kunder2012,Ness2013a}. This is indicative of formation from the disk via internal evolution \citep{Ness2016b,Fragkoudi2017,Buck2018}. However, the kinematic character of the bulge changes as a function of the stellar metallicity \citep[e.g.][]{Portail2016}. More metal rich stars show lower velocity dispersions \citep{Ness2013b} and the most metal-poor stars ($<$ --1.0) are nearly \textbf{slow-rotating} \citep{Kunder2016}. The origin of the metallicity dependent kinematics for stars that are in the boxy/peanut population has been interpreted as a consequence of a different mapping of stars into the peanut structure at onset of the bar instability. Thereby stars of the thin and the thick disk are mapped differently into the boxy/peanut structure \citep{Fragkoudi2017} depending on their initial kinematics \citep{Debattista2016}. Hereby, thin disk stars are characterised by their small in-plane velocity dispersions, which make them more susceptible to become part of the peanut structure. Conversely, thick disk stars are characterised by their large in-plane random motions. These stars become a thicker structure at the onset of the bar instability and trace out less of a less peanut/bar shape compared to the thin-disk stars. This effect was termed \textit{kinematic fractionation} by \citep{Debattista2016}. The metal-poorer stars ([Fe/H]$< −1.0$ dex) that are very slowly rotating have been associated with a population that is not part of the boxy bulge structure. These metal poor stars are possibly (a combination of) spherical bulge \citep[e.g.][]{DiMatteo2014,DiMatteo2015}, stellar halo \citep{Howes2016} and/or classical bulge, with an origin distinct from the disk \citep[e.g.][]{Rojas-Arriagada2017}.  However, the detailed formation of the metal-poor, slowly rotating population that is seen in the inner region of the MW and its formation timescale in relation to the boxy/peanut structure is not yet understood. 

Controlled $N$-body experiments \citep[e.g.][]{Athanassoula2009,DiMatteo2016,Athanassoula2017} are well suited to study the internal mechanisms at play in shaping the bulge of the MW. However, they exclude a fundamental ingredient necessary to capture the complex build-up of the central Galactic region. That is, that galaxies grow in a cosmological environment. Galaxies accrete gas, form stars, get bombarded and disturbed by satellite galaxies and self-enrich their gas with metals from stellar feedback. Including these additional processes in addition to the internally driven evolution is necessary in order to study the different components of galactic bulges and explain their origin. 

Only cosmological simulations are able to self-consistently capture these effects. However, for a long time cosmological simulations were not able to reproduce realistic bulges due to numerical resolution and the inherent hierarchical nature of galaxy formation. Cosmological simulations commonly predict that galactic spheroids are primarily built up through hierarchical mergers \citep[e.g.][but see also \citet{Obreja2018b}]{Kauffmann1993,Abadi2003,Kobayashi2011,Guedes2013}. This formation scenario produces an old classical bulge in contrast to a boxy/peanut shaped bulge such as is observed in the MW \citep{Weiland1994,Dwek1995,Ciambur2017}. Recently cosmological simulations have succeeded in reproducing barred spiral galaxies \citep[e.g.][]{Spinoso2017}. In \citet{Buck2018} (paper I hereafter) we have presented one of the first cosmological simulations able to reproduce several of the key characteristics of the MW central region including a realistic boxy/peanut morphology \citep[see also][]{Debattista2018}.

Given we can now reproduce MW type bulges in cosmological simulations, we have the opportunity to test the orbital structure of these objects in simulations \citep[see e.g.][]{Portail2015a,Portail2015}.
For the MW, we have the expectation that several stellar constituents co-exist in the inner region. Using an orbital decomposition of stars is a promising framework to understand bulge substructure and its formation origin, particularly in the Gaia era. 
From a dynamical point of view, stellar orbit classification provides the most robust way to disentangle different galactic stellar structures \citep{Binney2013}. Observationally, this approach can be applied to the MW, thanks to recent large scale surveys \citep[e.g.][]{Steinmetz2006,Brown2016,Majewski2017}. Stellar orbit classification has also been recently applied to extragalactic observations \citep{Zhu2017,Zhu2017a,Gonzalez2016,Gonzalez2017} of CALIFA galaxies \citep{Sanchez2012}. 

In this paper we use an unsupervised Gaussian mixtures algorithm in a particular stellar kinematic space \citep{Obreja2016,Obreja2018} in order to understand the orbital composition of the MW-like bulge in the high resolution galaxy of paper I, which has a clear boxy-peanut morphology and a cylindrical kinematic profile. Standard analysis of abundance patterns and line-of-sight velocities in the central regions of this galaxy suggest that large parts of the bulge and its central bar are formed from disk material (see paper I). Stellar particles of this simulated galaxy can be decompose into a high- and low-spin disk, a high- and low-spin bulge and a stellar halo population. In this paper we focus on the two bulge components and study in detail their properties and formation scenarios in order to investigate the driving mechanisms in separating stellar particles into different orbit families.

%%%%%%%%%%%%%%%%%% FIGURE 2 %%%%%%%%%%%%%%%%%%%%%%%%%%%
\begin{figure*}
\includegraphics[width=\textwidth]{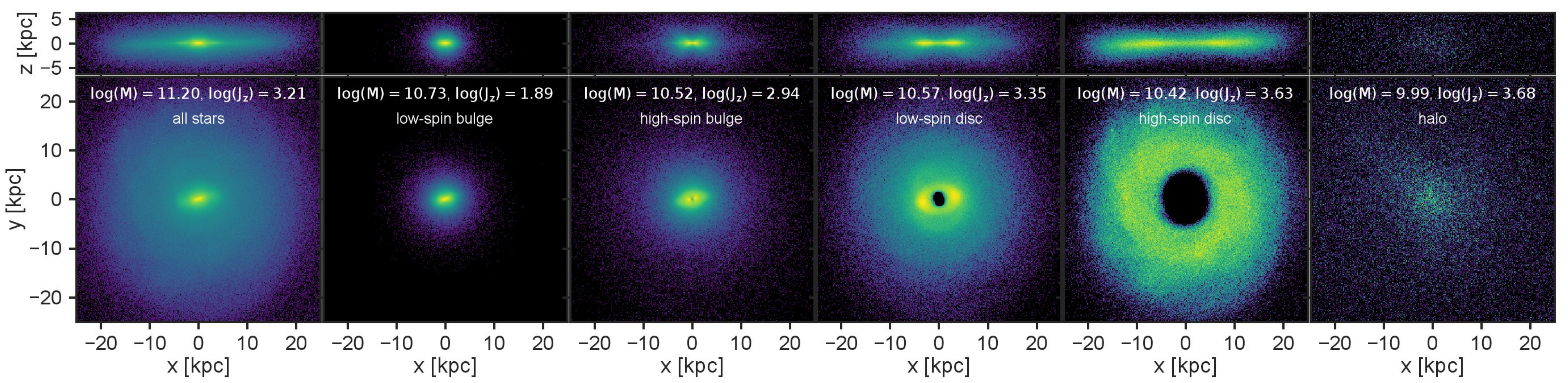}
%\vspace{-.25cm}
\caption{Surface density maps in edge-on (upper panels) and face-on (lower panels) view of all stars in the simulated galaxy (left most panels) and in the different kinematic components identified using the method of \citet{Obreja2018}. From left to right: all stars, low-spin bulge, high-spin bulge, low-spin disk, high-spin disk and stellar halo. Each panel indicates the total mass in M$_{\rm\odot}$ and the total specific angular momentum of each component in kpc~km~s$^{\rm-1}$.}
\label{fig:decomp_surf}
\end{figure*}
%%%%%%%%%%%%%%%%%%%%%%%%%%%%%%%%%%%%%%%%%%%%%%%%%%%%
 
The paper is structured as follows: In \S2 we briefly describe the simulation. In \S3 we introduce the decomposition technique, and in \S4 we discus the properties of the decomposed components. Then in \S5 we study in detail the kinematical properties of the bulge components in this simulation. In \S6 we analyse the origin of stellar populations in the bulge of the simulation and shed some light on why stellar particles end up in the central structure. In \S7 we discuss our results with respect to observational findings and take a closer look at the metallicity distribution of the bulge components. Finally, in \S8 we summarize our results and present our conclusions.

%%%%%%%%%%%%%%%%%%%%%%%%%%%%%%%%%%%%%%%%%%%%%%%%%%%
\section{Simulation and galaxy properties} \label{sec:simulation}
%%%%%%%%%%%%%%%%%%%%%%%%%%%%%%%%%%%%%%%%%%%%%%%%%%%

The simulation used in this work is a higher-resolution version of the galaxy g2.79e12 taken from the Numerical Investigation of a Hundred Astronomical Objects (NIHAO) project \citep{Wang2015}. This galaxy has been previously studied in paper I, where we showed it has a strong, long-lived bar. While in paper I we focused on an observationally motivated comparison between the properties of the high-spin bulge in the simulation and in the MW (e.g. by means of comparing simulated and measured rotation and dispersion profiles), this paper studies the properties of the central regions in a more theoretically motivated manner. The aim of this kind of analysis is to study properties and formation mechanisms of kinematically distinct components in order to facilitate upcoming interpretations from large scale Galactic surveys like Gaia \citep{Katz2018} or 4MOST \citep{DeJong2016}.

The hydrodynamics, star formation recipes and feedback schemes exploited are the same as for the original NIHAO runs and we refer the reader to the original NIHAO paper \citep{Wang2015} and paper I for more details. For completeness we briefly summarize key simulation parameters below.

This high-resolution simulation was run with a modified version of the smoothed particle hydrodynamics (SPH) code {\texttt{GASOLINE2}} \citep{Wadsley2017} using cosmological parameters from \cite{Planck}. Gas cooling is implemented via Hydrogen, Helium, and various  metal-lines as described  in \cite{Shen2010}, and cooling functions are calculated using \texttt{cloudy} \citep[version 07.02;][]{Ferland1998}. Star formation is implemented following \cite{Stinson2006} and two modes of stellar feedback are employed \citep{Stinson2013}. The free parameters in the feedback scheme are calibrated at MW mass against the stellar mass-halo mass relation at $z = 0$ \citep{Moster2013}.

The feedback scheme chosen for the NIHAO sample has been shown to match remarkably well many properties of observed galaxies. NIHAO galaxies are able to recover the local velocity function \citep{Maccio2016}, match the metal distribution in the circumgalactic medium \citep{Gutcke2016}, the properties of stellar disks \citep{Obreja2016} and gaseous disk \citep{Dutton2016b} as well as the morphological properties of high-mass galaxies at high redshift \citep{Buck2017}.

The mass resolution is $m_{\rm dark}=5.14\times10^5 \Msun$ for dark matter particles, $m_{\rm gas}=9.38\times10^4 \Msun$ for gas particles and the initial star particle mass is $m_{\rm star}=3.13\times 10^4\Msun$. The corresponding force softenings are $\epsilon_{\rm dark}=620$ pc and $\epsilon_{\rm gas}=\epsilon_{\rm star}=265$ pc for the gas and star particles. The final total mass within the virial radius (R$_{\rm vir}\sim 300$ kpc) is  M$_{\rm tot}=3.13\times10^{12} M_{\odot}$, the mass of the dark matter halo is M$_{\rm dark}=2.78\times10^{12} M_{\odot}$ and the stellar mass of the galaxy (measured within $0.1\times$R$_{\rm vir}$) is M$_{\rm star}=1.42\times10^{11} M_{\odot}$. The galaxy's stellar disk has a scale length of R$_{\rm d}\sim 5$ kpc and a scale height of H$_{z}\sim500$ pc within the innermost 5 kpc. As measured in paper I, the bar in this simulation formed around 8 Gyr ago at a redshift of $z\sim 1$. Such an early formation and persistence of a bar has recently been established observationally for the galaxy NGC 4371 \citep{Gadotti2015}. An all-sky view of the galaxy from the sun's position at 8 kpc from the center is given in Fig. \ref{fig:all_sky}.

%%%%%%%%%%%%%%%%%%%%%%%%%%%%%%%%%%%%%%%%%%%%%%%%%%%
\section{Stellar kinematics decomposition}
\label{sec:decomp}
%%%%%%%%%%%%%%%%%%%%%%%%%%%%%%%%%%%%%%%%%%%%%%%%%%%
\label{sec:bulge}

The current standard kinematic decomposition of simulated galaxies was first introduced by \cite{Abadi2003b} and relies on analyzing the distribution of stellar circularities, $\epsilon(E)=J_{\rm z}/J{\rm c}(E)$, which is computed from the azimuthal angular momentum of a particle,  $J_{\rm z}$, and the angular momentum of a circular orbit, $J_{\rm c}(E)$, having the same binding energy, $E$ \citep[e.g.][]{Brook2004,Brooks2008,Tissera2012}.
This decomposition method has widely been used to study disks and bulges in galaxy formation \citep[see e.g.][]{Scannapieco2010,Scannapieco2011,Martig2012,Marinacci2014,Kannan2015,Zavala2016}. However, this classical method makes ambiguous classifications of stars with intermediate values of circularities. In order to circumvent this ambiguity \cite{Domenech-Moral2012} suggested adding the binding energy, $E$, and the angular momentum component perpendicular to the disk, $J_{\rm p}$, to the parameter set. Furthermore, instead of adopting fixed values in the various parameters to diskriminate between disk and bulge, \cite{Domenech-Moral2012} used a cluster finding algorithms \citep[e.g. \emph{k-means}][]{kmeans} to arrive at data driven cuts. 

\citet{Obreja2016,Obreja2018} used the method outlined by \citet{Domenech-Moral2012}, but with the \emph{Gaussian Mixture Models} (GMM) of the machine learning code scikit-learn \citep{scikit-learn} as clustering algorithm instead of \emph{k-means}. The main reason for using GMM instead of \emph{k-means} is that the latter algorithm naturally results in clusters of roughly equal sizes, and as such it is not optimal for finding sub-dominant structures like stellar halos. The more recent method operates over a similar parameter space as suggested by \cite{Domenech-Moral2012}. However, to factor out the galaxy/halo mass dependence and to remove the dimensionality of the energy, \cite{Obreja2016} switched to specific values of angular momentum and binding energy, $(j_{\rm z}/j_{\rm c}, j_{\rm p}/j_{\rm c}, e/\rm{max}(|e|))$. Additionally, the binding energies of stellar particles are scaled to fall within -1 to 0, such that -1 represents the most bound star particle. 

One advantage of using data-driven decomposition techniques such as GMM, is the flexibility in being able to define this number of components. Using GMM, the only input-parameter to be provided is the number of components. We can also determine, by iterating over the number of components,  how many components best fit the data. For this study we started with a 2 component specification (disk- and spheroidal-component) and increased the number of components successively. We found that a total of 5 different components best represents the data in our simulation. Fig.~\ref{fig:decomp_distr} in the Appendix gives the normalized stellar mass distributions of all 5 components in the input kinematic space of $(j_{\rm z}/j_{\rm c}, j_{\rm p}/j_{\rm c}, e/\rm{max}(|e|))$. 
The choice of five components is based on the visual inspection of the maps of edge-on surface mass density and line-of-sight velocity. The first components to be well separated by GMM are the disk(s), hence we kept on increasing the number of components to asses weather the galaxy has one or two (thin and thick) extended disks, and whether the algorithm can separate the stellar halo. We have also studied if various information criteria like the BIC \citep{Schwarz1978} or AIC \citep{Akaike1974} can be used for model selection in this context using the galaxy sample presented in \citet{Obreja2018}. Unfortunately, none of these criteria turn out to be suitable for our problem, which is essentially an unsupervised classification one. Fig. \ref{fig:bic} of the Appendix shows how the negative log likelihood ($-\log\left(L\right)$) varies as a function of $k$.

%%%%%%%%%%%%%%%%%% FIGURE 3 %%%%%%%%%%%%%%%%%%%%%%%%%%%
\begin{figure*}
\includegraphics[width=.33\textwidth]{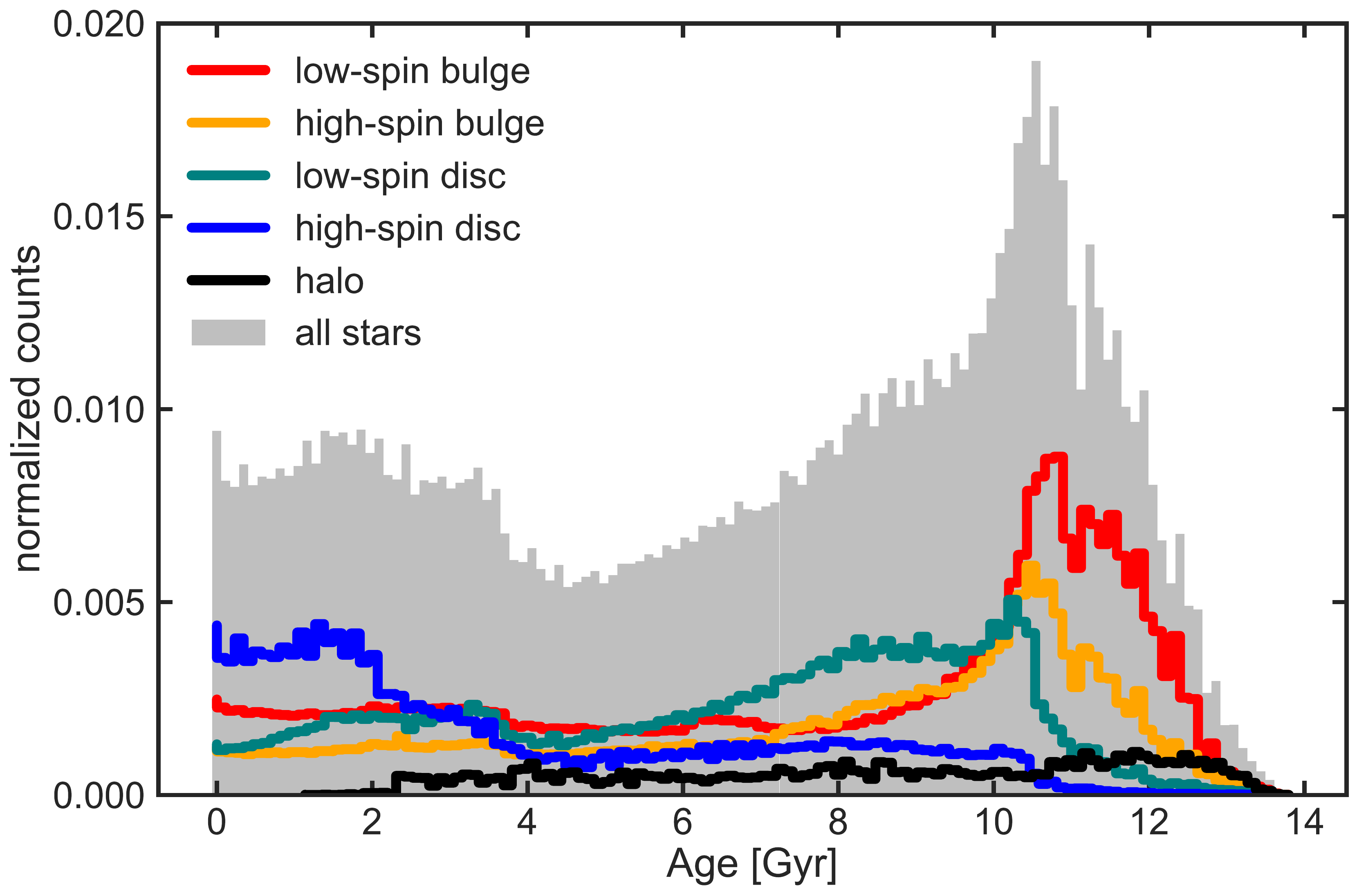}
\includegraphics[width=.33\textwidth]{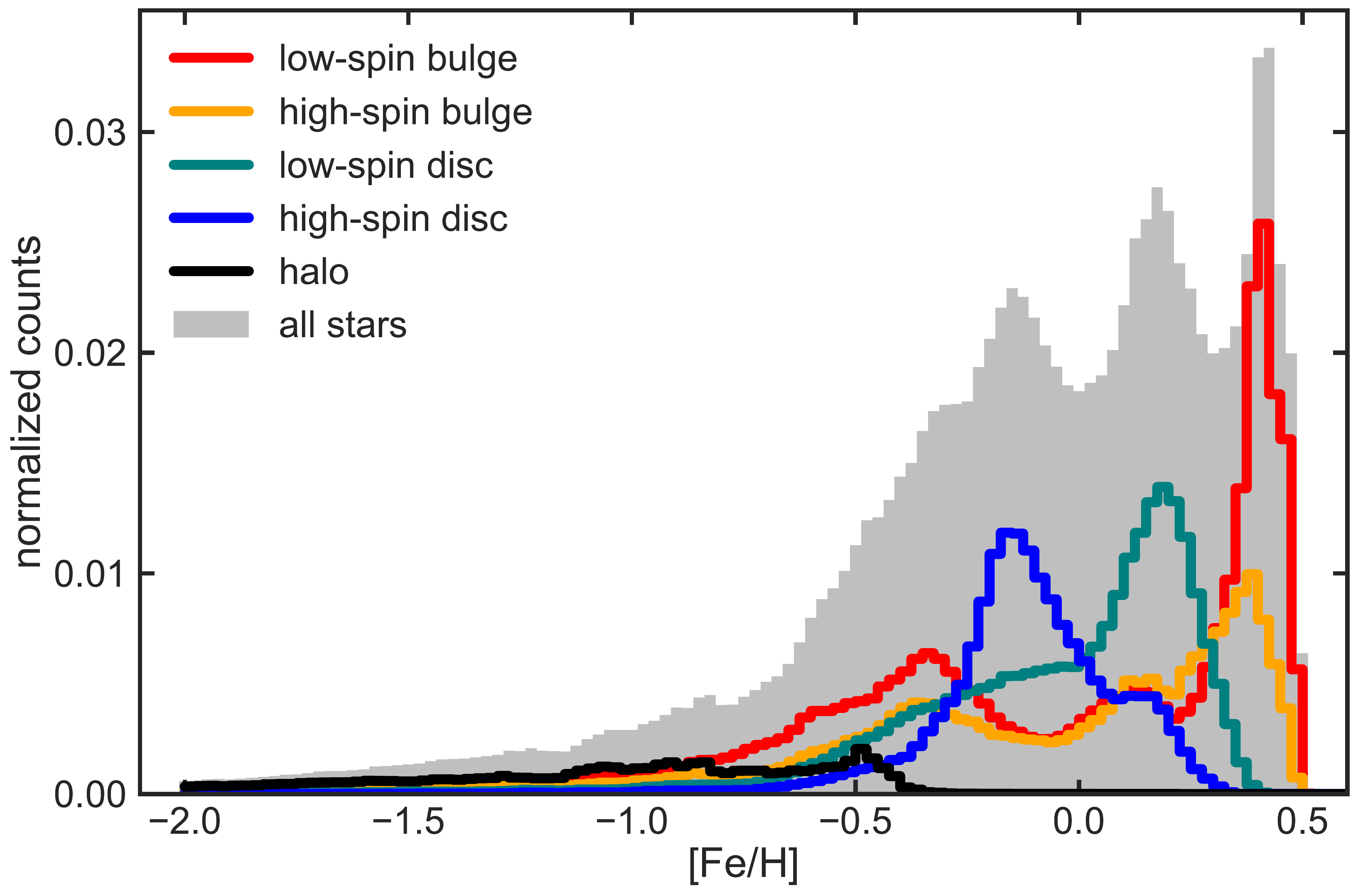}
\includegraphics[width=.32\textwidth]{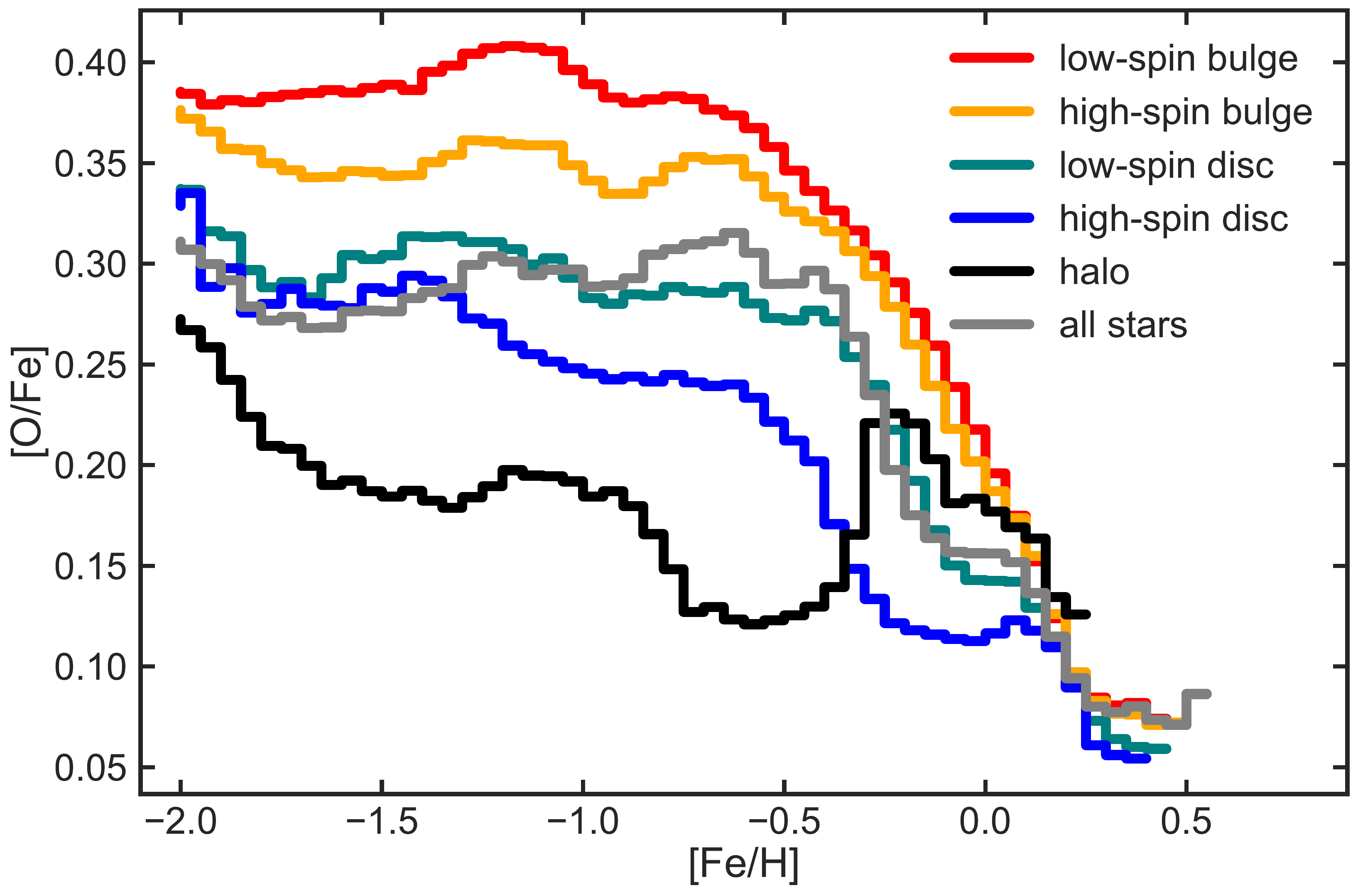}
%\vspace{-.25cm}
\caption{Properties of the kinematically decomposed populations of stars. From left to right we show the mass-weighted age distribution, the metallicity distribution and the oxygen abundance [O/Fe] as a proxy for $\alpha$-elements vs. metallicity. Colored lines show the different components as indicated in the legend and the gray filled histogram or the gray thick line (right panel) shows the distribution for the whole stellar population of the galaxy. The scatter around the mean is much smaller than the line thickness for the two left most panels and about $\Delta$[O/Fe]$\sim0.05$ at low metallicities and $\Delta$[O/Fe]$\sim0.025$ at the high metallicity end.
}
\label{fig:decomp_props}
\end{figure*}
%%%%%%%%%%%%%%%%%%%%%%%%%%%%%%%%%%%%%%%%%%%%%%%%%%%%

\section{Kinematic components properties}
\subsection{Morphology and angular momentum}

Figure \ref{fig:decomp_surf} shows the surface density maps in edge-on and face-on projections for all the stars in the galaxy (left most column) and for the five different components identified in the subsequent columns. We stress that we define these five components \textbf{solely} via their kinematics and not their morphological properties (see also Fig. \ref{fig:decomp_distr} in the appendix for the distributions of j$_{\rm z}$/j$_{\rm c}$, j$_{\rm p}$/j$_{\rm c}$, and e/max($\vert$e$\vert$) for each component).

We name the components returned by the algorithm based on their total angular momentum as follows: a low-spin bulge component, a high-spin bulge\footnote{There is some confusion in the literature about the usage of peanut or X-shaped bulge and bar. Here we use both terms as synonyms for the same component as the peanut shaped bulge or equivalently the high-spin bulge of our simulation is essentially the buckled bar.}, a low-spin disk, a high-spin disk and a spheroidal stellar halo component (from left to right in Fig \ref{fig:decomp_surf}). These names are chosen to foremost describe the kinematical properties of the components. However, when looking at the overall morphology of these components we can see that the high spin disk indeed is morphologically thin while the low-spin disk is morphologically thick. On the other hand, the low-spin bulge component shows a spherically symmetric morphology and the high-spin bulge is of boxy/peanut shape. We do not claim these necessarily reflect the specific individual populations seen in the MW and leave a detailed comparison of the similarities and dissimilarities (e.g. alpha enhancement of the low-spin disk) of the disk components with observationally identified disk components in the MW for future work. In each panel, we indicate the stellar mass in each component and its specific angular momentum. Except for the stellar halo component, all others contribute nearly an equal amount of stellar mass to the whole stellar budget of the galaxy. Interestingly, the two disk components reveal holes in the central region. This implies that all stellar particles in this region are on hot orbits and are thus assigned to the other three components - the two bulge components and the halo \citep[see also][]{Portail2017}. 

The low-spin bulge component shows a spherically symmetric morphology. The high-spin bulge component is barred in face-on view and shows signs of an X-shaped morphology in edge-on view, with a very thin component extending out to $\sim10$ kpc in the disk mid-plane. The low-spin bulge component shows the lowest value of total angular momentum, indicating almost no net rotation. The high-spin bulge component has a much larger total angular momentum. The low-spin disk component is short and extends up to heights of $\sim4$ kpc above the disk mid-plane. The high-spin disk component is much more extended, and appears thinner in the edge-on view. The low-spin disk component shows a lower value of total angular momentum than the high-spin disk, but a larger value compared to the high-spin bulge. The largest value of total angular momentum is found for the stellar halo component. This halo component is by far the smallest component in terms of mass. We attribute the high angular momentum amplitude to its much more extended morphology. 

\subsection{Stellar ages}
\label{sec:age_props}

The left panel of Fig. \ref{fig:decomp_props} shows the mass-weighted distribution of stellar ages in each component (colored lines) as well as in the whole galaxy (gray filled histogram).

We find that components differ in their median age, although stars are distributed across a broad age range for all of them. The low-spin bulge (red line) has the largest fraction of old stars. The age distribution peaks around 10-13 Gyr, at higher ages compared to all other components. The high-spin bulge (yellow line) is the second-oldest component overall, with the peak shifted to slightly lower stellar ages, of around 10 Gyr. Interestingly, both bulge components show tails down to the youngest stellar ages. We shall come back to this point in the discussion. The low-spin disk (green line) shows a broad peak around stellar ages of $\sim$8 Gyr, and the high-spin disk (blue line), by far the youngest of the five, shows a clear peak at stellar ages of about 1 Gyr, with a tail towards older stellar ages. The stellar halo has a far flatter age distribution and exhibits stars of all ages down to 2.5 Gyr (there are no very young stars in the halo with ages $<$ 2.5 Gyr).

\subsection{Stellar metallicities}
\label{sec:met_props}
% metallicities
The components also separate out in the median and peak of their metallicity distributions, as shown in the middle panel of Fig. \ref{fig:decomp_props}. The low-spin bulge, seen to be the oldest component in the left panel of Figure \ref{fig:decomp_props}, shows a strong peak at the highest metallicities, of values above solar ($\sim0.4$) and smaller peaks at around solar metallicity and $\sim-0.4$ and below (red line). The high-spin bulge (yellow line), the second oldest component,  has a similar metallicity distribution to the low-spin bulge, but shifted to slightly lower metallicities.  The low-spin disk (green line), shows a strong and broad peak around metallicities of about 0 to 0.2 dex with a tail down to -0.5 dex while the high-spin disk (blue line) shows a well defined single peak around solar metallicity. We find that the low-spin disk exhibits slightly more stars at higher metallicities compared to the high-spin disk. Finally, the stellar halo (gray line) is made up of stars of the lowest metallicities and shows (only) a broad distribution below metallicities of -0.4 dex. The total metallicity distribution function (black line) shows three distinct peaks at slightly sub-solar ($\sim-0.2$ dex), slightly super-solar values ($\sim0.2$ dex) and 0.4 dex. Comparing the single components to the total metallicity distribution, we see that these peaks are made up of primarily the high-spin disk, low-spin disk ,and bulge stars.

\subsubsection{[O/Fe] vs. [Fe/H]}
%  [O/Fe] vs. [Fe/H] distribution
In the right panel of Fig. \ref{fig:decomp_props}, we show the average [O/Fe] vs. [Fe/H] distribution for all five sub-components (thick colored lines) with the same line colors as before. We recover the expected trend in oxygen enrichment as a function of metallicity. The low-spin bulge is the most $\alpha$-enhanced population, followed by the high-spin bulge. Low- and high-spin disk separate in the metallicity range of -1.0 dex to 0.0 dex, with the high-spin disk being offset to lower values of [O/Fe] (interestingly, these show similar $\alpha$-enhancement at very low $<$ (-1.0 dex) and very high ($>$ 0 dex) metallicities). Comparing this finding with the metallicity distribution function of the two disks we conclude that the high metallicity part of the low-spin disk might come from thin disk stars which get kinematically heated. By far the lowest values of [O/Fe] are found for the stellar particles in the halo component. In this panel, the gray thick line shows the result using all stellar particles in the disk.

The MW shows two alpha sequences, a low and a high sequence \citep[e.g. see][]{Hayden2015} that have been associated with a thin and thick disk component \citep[see e.g.][]{Gilmore1983}. We highlight again here that we recover a difference in the $\alpha$-enhancement of the high- and low-spin disk star particles, using a decomposition solely based on stellar kinematics. This result is worth investigating in more detail in subsequent work, but is beyond the scope of this paper. 

%%%%%%%%%%%%%%%%%% FIGURE 7 %%%%%%%%%%%%%%%%%%%%%%%%%%%
\begin{figure*}
\includegraphics[width=\textwidth]{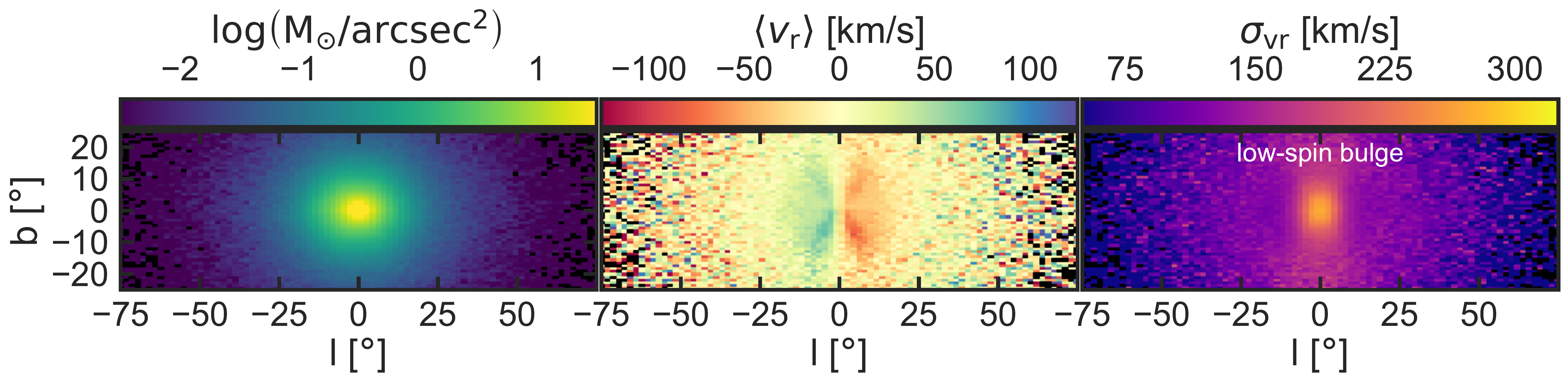}
\includegraphics[width=\textwidth]{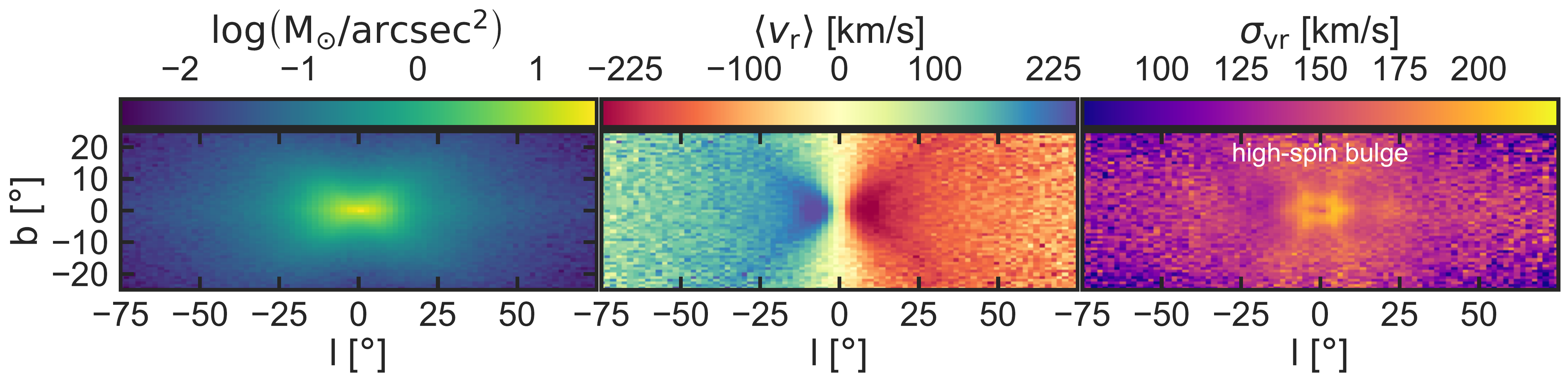}
\vspace{-.35cm}
\caption{Surface density (left column), mass-weighted rotation (middle column) and dispersion maps (right column) in $(l,b)$-projection. for the . The top row shows the results for the low-spin bulge component and the bottom row for the high-spin bulge component respectively.
}
\label{fig:cb_rot_disp_map}
\end{figure*}
%%%%%%%%%%%%%%%%%%%%%%%%%%%%%%%%%%%%%%%%%%%%%%%%%%%%

\section{Properties of the ``low-spin bulge" and the ``high-spin bulge"}
\label{sec:props}

In order to compare properties of the two bulge components with observations of the MW, we rotate our galaxy to a heliocentric frame of reference where the stellar disk is in the $x-y$ plane and the bar inclined at $27^\circ$ with respect to the line-of-sight. We then observe the bulge from the sun's position at a distance of $8$ kpc from the galactic center (see also paper I for more details).

\subsection{Rotation and dispersion maps} 

In Fig.\ref{fig:cb_rot_disp_map}, we show maps of the stellar surface density (left column), rotation (middle column) and dispersion values (right column) of the low-spin and the high-spin bulge in $l,b$. In the Appendix we show a similar figure for all kinematically identified components (see Fig. \ref{fig:maps_decomp}).  
% rotation

The surface density maps clearly reveal the different morphologies for the low-spin and the high-spin bulges. They further show that the low-spin bulge component shows only small to no rotation, while the high-spin bulge component shows a fast rotation. The clear X-shaped structure in the rotation map is indicative of bar/box orbits viewed edge-on.

%dispersion
The  dispersion map of the low-spin bulge component shows remarkable spherical symmetry with a strong peak in the center, and a very latitude dependent dispersion profile. On the contrary, the dispersion map of the high-spin bulge component shows strong features due to the underlying morphology. The map shows a clear X-shaped dispersion signature in the center, resembling the morphology of the X in the surface density projections. Apart from this, the dispersion map is relatively flat and does not show an increase in dispersion towards the center, as seen for the low-spin bulge component. Thus, dissimilarly to the low-spin bulge component, the high-spin bulge component has a fairly latitude independent dispersion. 

\subsection{Rotation and dispersion profiles compared to observations}

\begin{figure*}
\centering
\includegraphics[width=.75\textwidth]{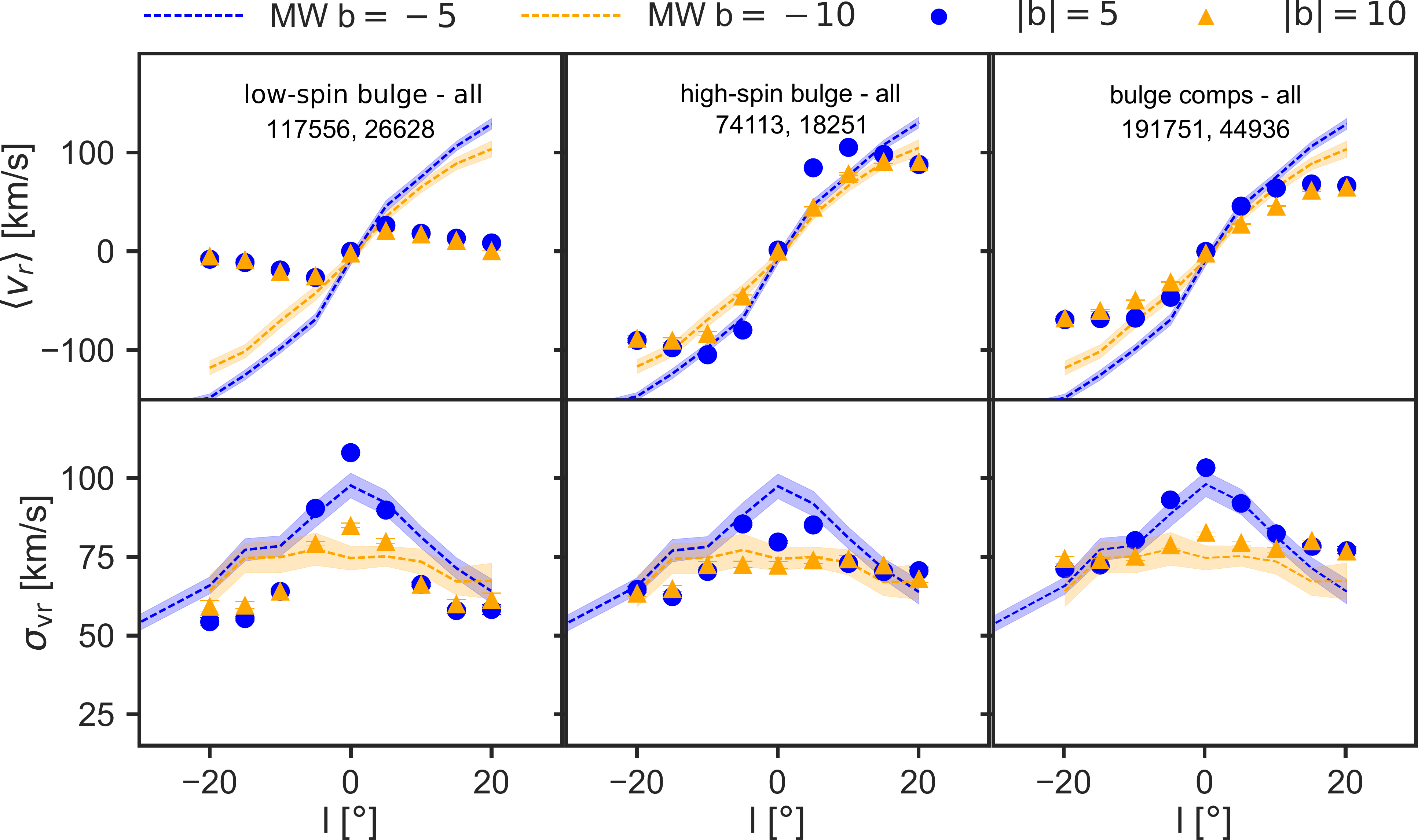}
%\vspace{-.35cm}
\caption{Rotation (upper row) and dispersion profiles (bottom row) for stars of the low-spin bulge (left panel), the high-spin bulge (middle panel) and both components combined (right panel) for two different latitudes $\vert b\vert=5^\circ$ (blue) and $\vert b\vert=10^\circ$ (yellow). The rotation profile is the mean radial velocity along the line-of-sight and the dispersion profile is the velocity dispersion along the line-of-sight as a function of galactic longitude $l$. Lines show the measurements for the MW from the ARGOS survey \citep{Ness2013a,Ness2013b} and dots/triangles show the result obtained for our simulation. The numbers in each panel indicate the amount of stellar particles in each sample, the first number refers to sight lines with $\vert b\vert=5^\circ$, the second to $\vert b\vert=10^\circ$.
}
\label{fig:rot_disp}
\end{figure*}

The rotation and dispersion maps for the two bulge components shown in Figure \ref{fig:cb_rot_disp_map} highlight the different signatures we clearly see when breaking up the stars by their orbital classes. Observationally, we can not diskriminate orbital classes via line of sight velocities which have been measured for the bulge in pencil beam studies. We are therefore interested to see how the  kinematics of both bulge components individually and also their combination compare with the observations of bulge stars. Therefore we show in Fig. \ref{fig:rot_disp} the rotation (mean velocity along the line-of-sight, top row) and dispersion (velocity dispersion along the line-of-sight, bottom row) profiles, as a function of longitude for two different latitudes. The left panels show the stars of the low-spin bulge component, the middle panels the ones of the high-spin bulge and the right panels the combination of both components in comparison to observational data from the ARGOS survey \citep[colored dashed lines][]{Ness2013a}. Blue dots and yellow triangles dots show the profiles for latitudes of $b=5^\circ$ and $b=10^\circ$, respectively. As shown in paper I the model galaxy has higher total stellar mass compared to the MW, thus the overall rotation of stars is higher. In order to account for the stellar mass difference between the MW and our simulation we apply the same rescaling of all the velocities as done in paper I. This rescaling is valid since we are here only interested in the shape of the rotation and dispersion profiles and not their absolute values.

%%%%%%%%%%%%%%%%%% FIGURE 6 %%%%%%%%%%%%%%%%%%%%%%%%%%%
%%%%%%%%%%%%%%%%%%%%%%%%%%%%%%%%%%%%%%%%%%%%%%%%%%%%

%comparison among the two components
%rotation
We start the discussion of Fig \ref{fig:rot_disp} by first comparing the rotation and dispersion profiles of the two bulge components to each other and then turn to compare them in detail to the observations. We find that the low-spin bulge shows at most slow ($<50$ km/s) rotation, while the high-spin bulge is rotating fast ($\sim100$ km/s). The rotation profile of the low-spin bulge is latitude independent while the high-spin bulge shows slower rotation at $\vert b\vert=10^\circ$ compared to the rotation speed at $\vert b\vert=5^\circ$ which shows slight deviation from cylindrical rotation. Furthermore, its rotation profile is similar in shape and magnitude to the profile of the low-spin disk component (compare Fig. \ref{fig:rot_disp_decomp} in the Appendix). The profile rises quickly to its maximum value and then flattens, well in agreement with what is observed from the BRAVA survey \citep{Howard2008}.

%dispersion
The velocity dispersion profile of the low-spin bulge (lower panels in Fig. \ref{fig:rot_disp}) is triangular shaped for both latitudes shown and the peak at $l=0^\circ$ is stronger closer to the disk mid-plane at $\vert b\vert=5^\circ$ compared to $\vert b\vert=10^\circ$.
The dispersion profile of the high-spin bulge in turn is flatter and less triangularly shaped compared to the low-spin bulge. There is still a small peak in velocity dispersion in the center close to the disk at $\vert b\vert=5^\circ$\footnote{The dip in central velocity dispersion of the high-spin bulge components is most likely due to the missing stars in the very center.}, while for larger heights from the disk at $\vert b\vert=10^\circ$ the velocity dispersion profile is flat. This shows that the the high-spin bulge component is rotationally supported while the low-spin bulge is dispersion dominated as we have seen from Fig. \ref{fig:cb_rot_disp_map}.

%comparison of observed rotation profiles and simulated rotation profiles
We now compare the rotation and dispersion profiles of our simulation to observations from the ARGOS survey (dashed lines in Fig. \ref{fig:rot_disp}). We caution that this comparison of simulation and observation is only of qualitative nature because we compare a sub-population of stellar particles (low-spin bulge or high-spin bulge) to the full observational sample where such a distinction is not made. However, in doing so we can learn about the importance of an underlying low-spin bulge component in the MW. 

In comparison to the observed rotation profiles of the ARGOS survey, we see that the low-spin bulge component is not able to fully recover the observed rotation, although for longitudes in between $-5^\circ<l<5^\circ$ the profiles agree well. In contrast, the rotation profile of the high-spin bulge is in agreement with the observed profile, however, with slightly too large values of rotation close to the disk ($b=5^\circ$) at positive longitudes.

%comparison of observed dispersion profiles and simulated dispersion profiles
Comparing the dispersion profile of the low-spin bulge component to the observed profile we find that for both latitudes the simulated profile is far too peaked to be in agreement with the observations. This is somehow not surprising since we compare a single ``kinematically" selected sub-population in the simulation, the dispersion dominated population, to the full population in the observation with no further kinematical selection applied. 
The high-spin bulge fits the observed dispersion profile of the MW much better. It shows a peaked dispersion profile close to the disk mid-plane (blue dots) while it is flat for larger latitude (yellow dots).

% combination of spherical and \textbf{high-spin bulge}
Finally, if we do not distinguish the two bulge components and analyse the rotation and dispersion profile of the combination of stellar particles of both components (right panel of Fig. \ref{fig:rot_disp}) we find good agreement with the observations. We attribute the discrepancies at larger longitudes to the fact that we compare components that are centrally concentrated and end at $l\sim25^\circ$ to the full data of the MW which includes contributions from the disk at these large longitudes. This is similar to what we did in Paper I, Fig. 8. Thus, no single profile of either the low-spin bulge nor of the high-spin bulge is able to reproduce both the observed profiles exactly while their joint profiles give a better match to the MW data considering both the rotation and dispersion profile at the same time (especially if only considering the velocity dispersion profile). This makes us conclude, that the MW bulge might not be a pure high-spin nor a pure low-spin bulge but rather a mixture of both such components \cite[see also][]{Debattista2016}. Since the distinction of the two bulge components in the simulation is purely based on stellar kinematics, we predict that if with future surveys we are able to get orbital information for stars in the MW bulge we would be able to disentangle different kinematical components (e.g. low-spin vs. high-spin stellar distributions) in the bulge.

%We find that the oldest, most metal poor component of the stellar halo to be non-rotating in the inner region - halo, RR Lyrae, Kunder. 

\section{Origin of the components} 
 
\subsection{Origin of the two bulge populations}
\label{sec:birth}

%%%%%%%%%%%%%%%%%% FIGURE 10 %%%%%%%%%%%%%%%%%%%%%%%%%%%
\begin{figure}
\includegraphics[width=\columnwidth]{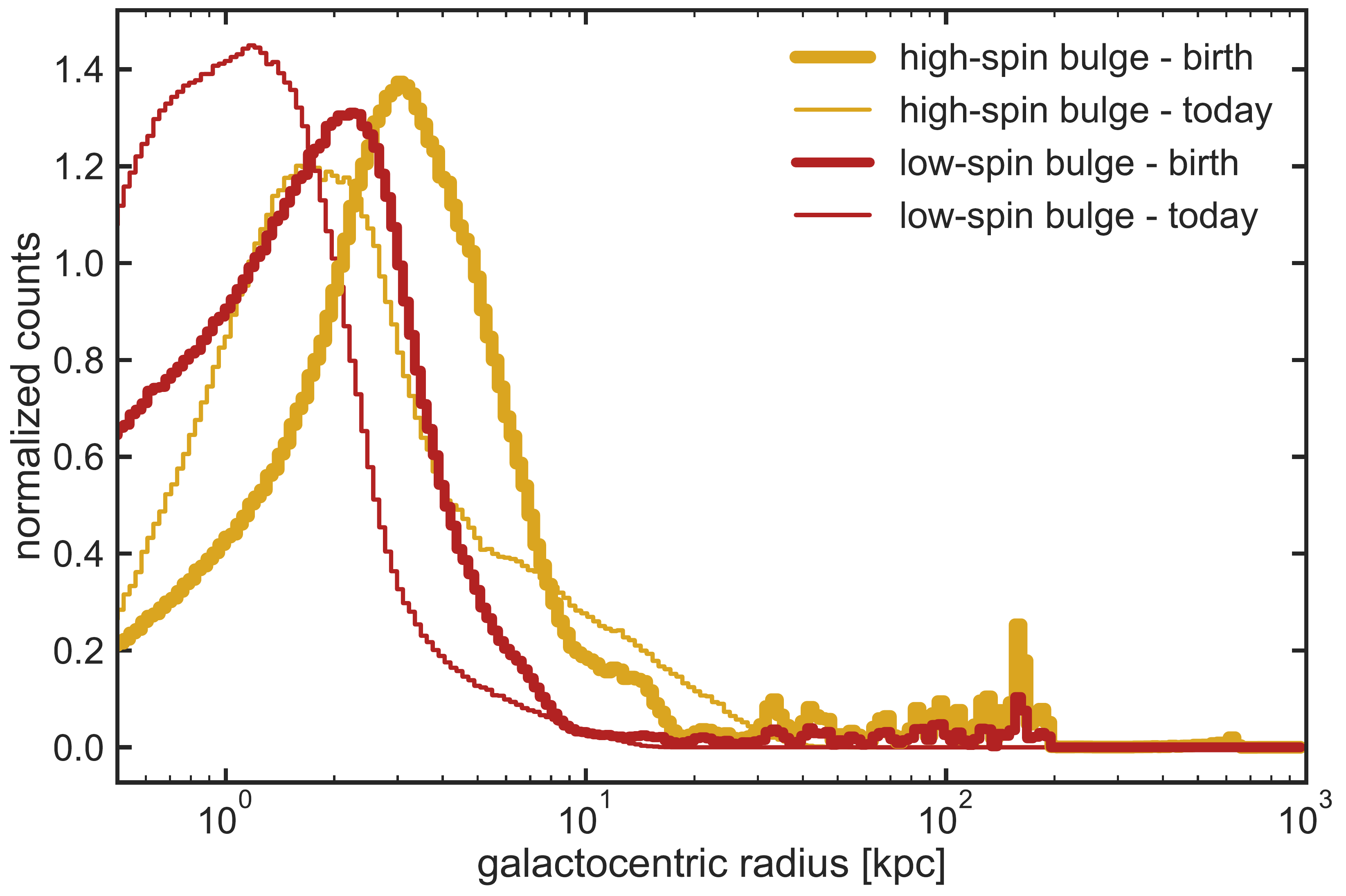}
\vspace{-.35cm}
\caption{Comparison between the birth radius (thick lines) of stellar particles and their present day radius (thin lines) with respect to the galaxy center for stellar particles in the low-spin bulge (red histogram) component and the high-spin bulge component (orange histogram).}
\label{fig:birth_pos}
\end{figure}
%%%%%%%%%%%%%%%%%%%%%%%%%%%%%%%%%%%%%%%%%%%%%%%%%%%%

The close similarity in chemical abundances (and stellar ages) for the low-spin and high-spin bulges implies a common origin (birth time). In fact, both components have a common disk origin as previously suggested by various authors \citep[e.g.][]{DiMatteo2015,Fragkoudi2017} but their present-day kinematical properties are distinct. The bar instability must have effected those stars in the high- and low-spin bulge differently to change their orbital properties. In order to investigate the reason for the separation of the two components, we trace back all stellar particles of each sub-component to their birth position, via their unique particle ID and record their birth radius and birth angular momentum.

\subsubsection{Birth position}
% general comparison, birth pos
In Fig. \ref{fig:birth_pos}, we compare the birth galactocentric radius (thick lines) to the present day galactocentric radius (thin lines) for the low-spin (red histogram) and the high-spin bulge (orange histogram). A figure comparing the birth height above the stellar disk is shown in the Appendix (see Fig. \ref{fig:birth_height}). The birth radius of the low-spin bulge component peaks at $\sim 2$ kpc and for the high-spin bulge component at $\sim 3$ kpc. There is only a slight difference between the overall birth radius distribution of stellar particles in the two components. Stellar particles belonging to the high-spin bulge show on average slightly larger (by about 1 kpc) present-day and birth radii compared to those of the low-spin bulge. The former result is not surprising, since we have already seen in Fig. \ref{fig:decomp_surf} that the high-spin bulge has a larger spatial extent.
For both components, the birth radius distribution peaks at larger radii compared to the present-day distribution. This indicates that material has been effected by the bar, leading to a redistribution of angular momentum such that stars originally on disk orbits loose angular momentum in $z$-direction and gain angular momentum perpendicular to the $z$-direction. This leads to larger excursions of the stars above the disk. This points towards a significant fraction of disk stars in these two components \citep[e.g.][]{DiMatteo2015,DiMatteo2016}. In fact, since both components show only mild contributions from stars born outside 100 kpc, which we refer to as ex situ stars, most of the stellar mass in the two bulge components originates from disk material. This disk material gets mapped into the bulge component by the bar, as previously pointed out by \citet[e.g.][]{Fragkoudi2017}. In terms of total stellar mass these ex-situ stars contribute about $0.7\%$. We investigate if and how these ex-situ stars are different from the rest of the stars in the next section.

Fig. \ref{fig:birth_pos} shows that populations of stellar particles selected to be on different orbits at the present day (low angular momentum, non-rotation for the low-spin bulge and higher angular momentum and rotation for the high-spin bulge) show distinct morphologies -- but are indeed born at fairly similar radii and times. Thus, we conclude that their birth position is not the main cause for different present day orbits. The next birth property to investigate is the birth angular momentum of stars. 

%%%%%%%%%%%%%%%%%% FIGURE 11 %%%%%%%%%%%%%%%%%%%%%%%%%%%
\begin{figure}
\includegraphics[width=\columnwidth]{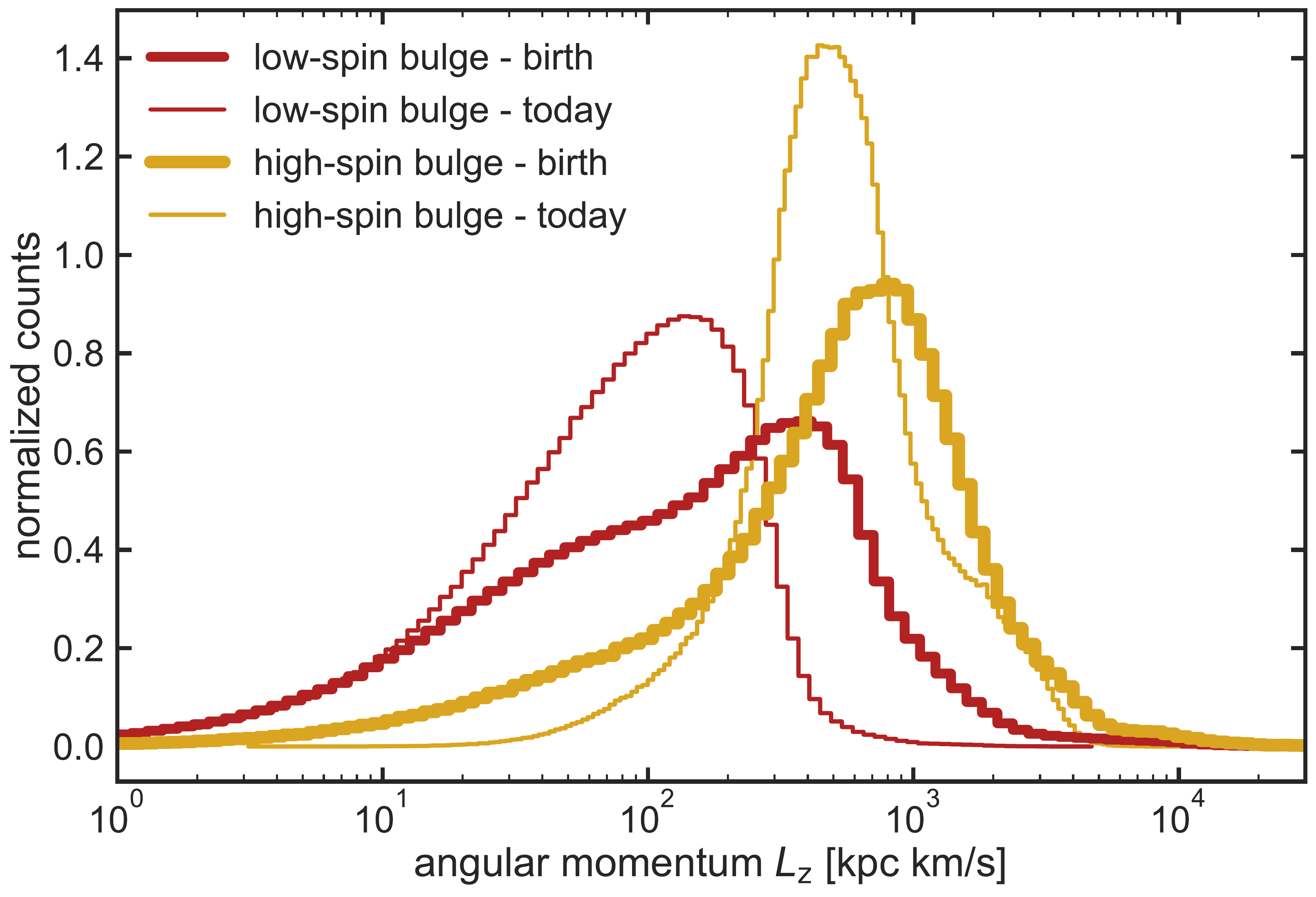}
\vspace{-.35cm}
\caption{Comparison between the birth angular momentum (thick lines) in $z$-direction and the present day angular momentum (thin lines) in $z$-direction (for the stellar disk in the $x-y$-plane) for stellar particles in the low-spin bulge (red histogram) component and the high-spin bulge component (orange histogram).}
\label{fig:birth_ang_mom_z}
\end{figure}
%%%%%%%%%%%%%%%%%%%%%%%%%%%%%%%%%%%%%%%%%%%%%%%%%%%%

%%%%%%%%%%%%%%%%%% FIGURE 12 %%%%%%%%%%%%%%%%%%%%%%%%%%%
\begin{figure*}
\includegraphics[width=.33\textwidth]{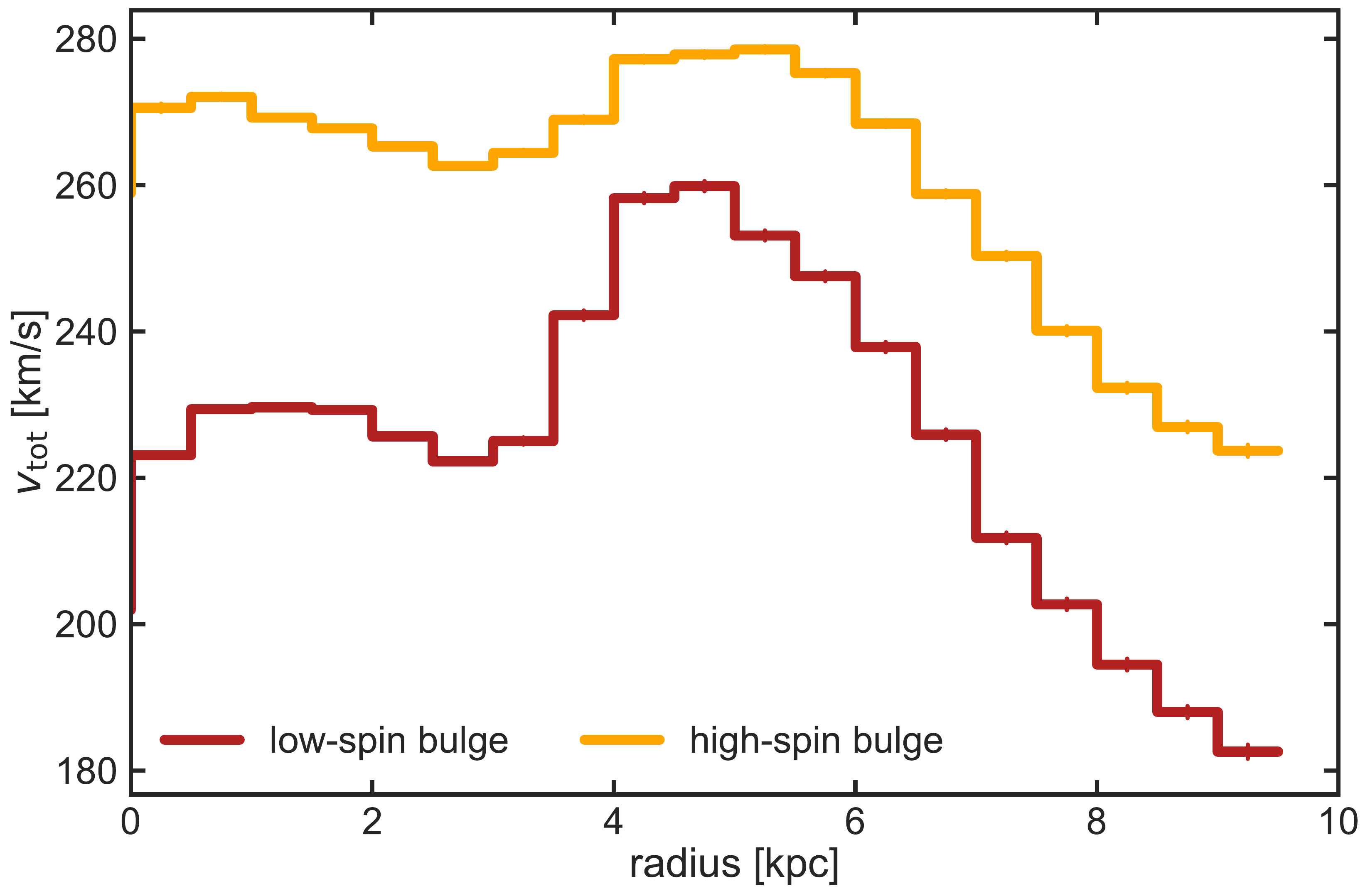}
\includegraphics[width=.33\textwidth]{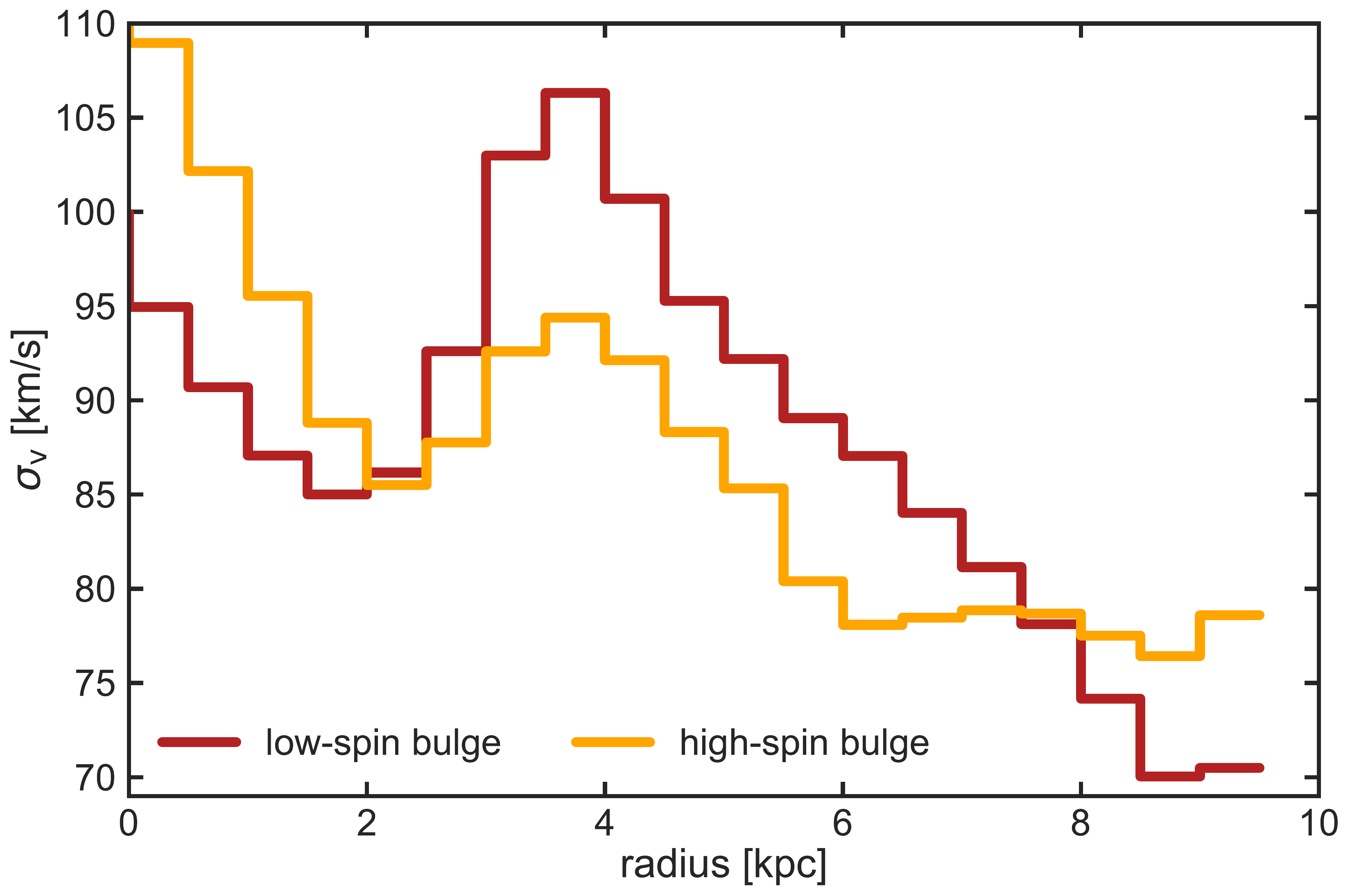}
\includegraphics[width=.33\textwidth]{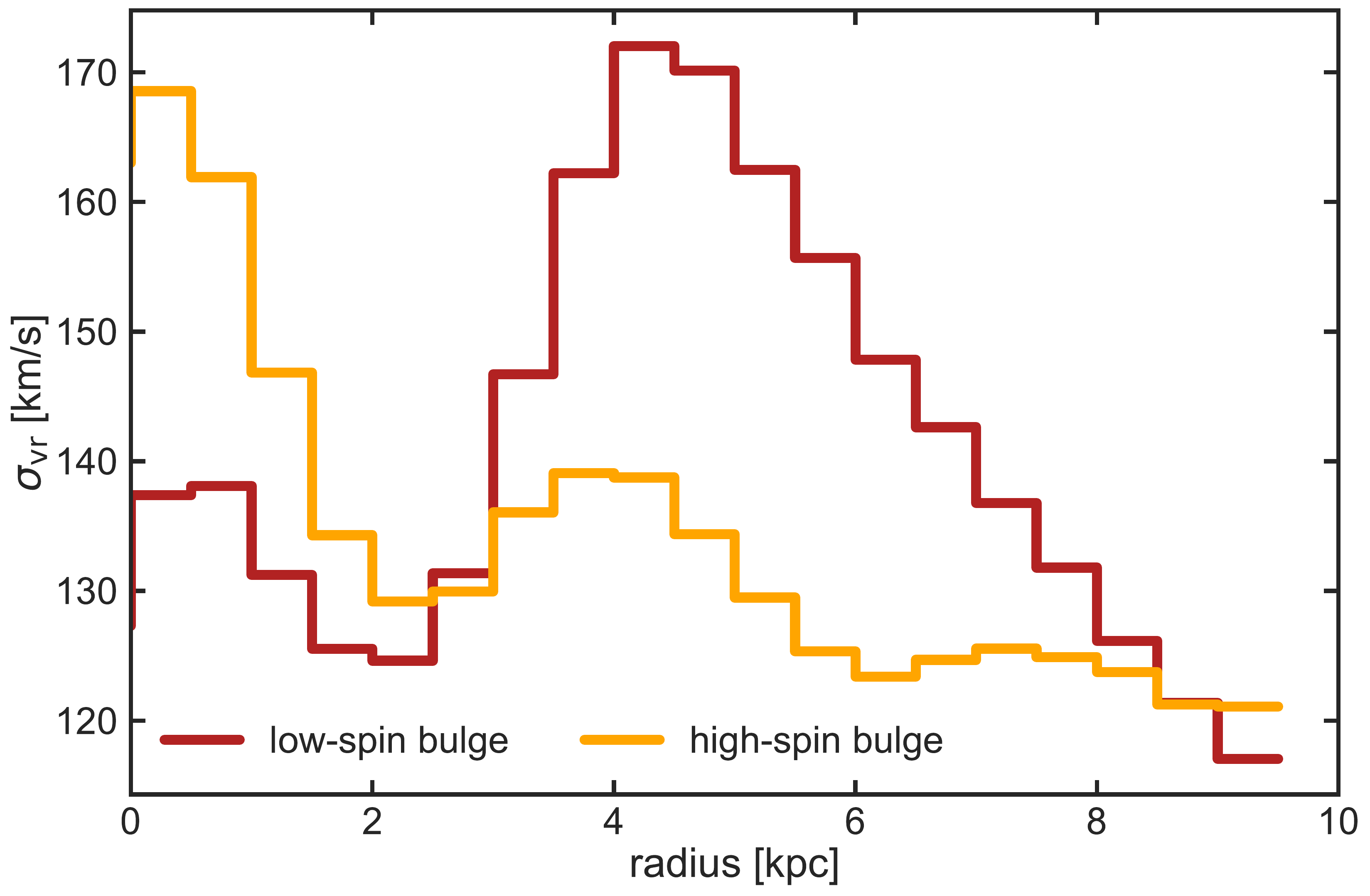}
\vspace{-.35cm}
\caption{Comparison between the total velocity (left panel), the total velocity dispersion (middle panel) and the radial velocity dispersion (right panel) for both bulge components shortly before bar formation at time $t\sim 5$ Gyr ($\sim8.7$ Gyr ago). Stellar particles in the low-spin bulge are shown in red and the high-spin bulge component is shown in orange.}
\label{fig:kin}
\end{figure*}
%%%%%%%%%%%%%%%%%%%%%%%%%%%%%%%%%%%%%%%%%%%%%%%%%%%%

%%%%%%%%%%%%%%%%%% FIGURE 13 %%%%%%%%%%%%%%%%%%%%%%%%%%%
\begin{figure*}
\includegraphics[width=.5\textwidth]{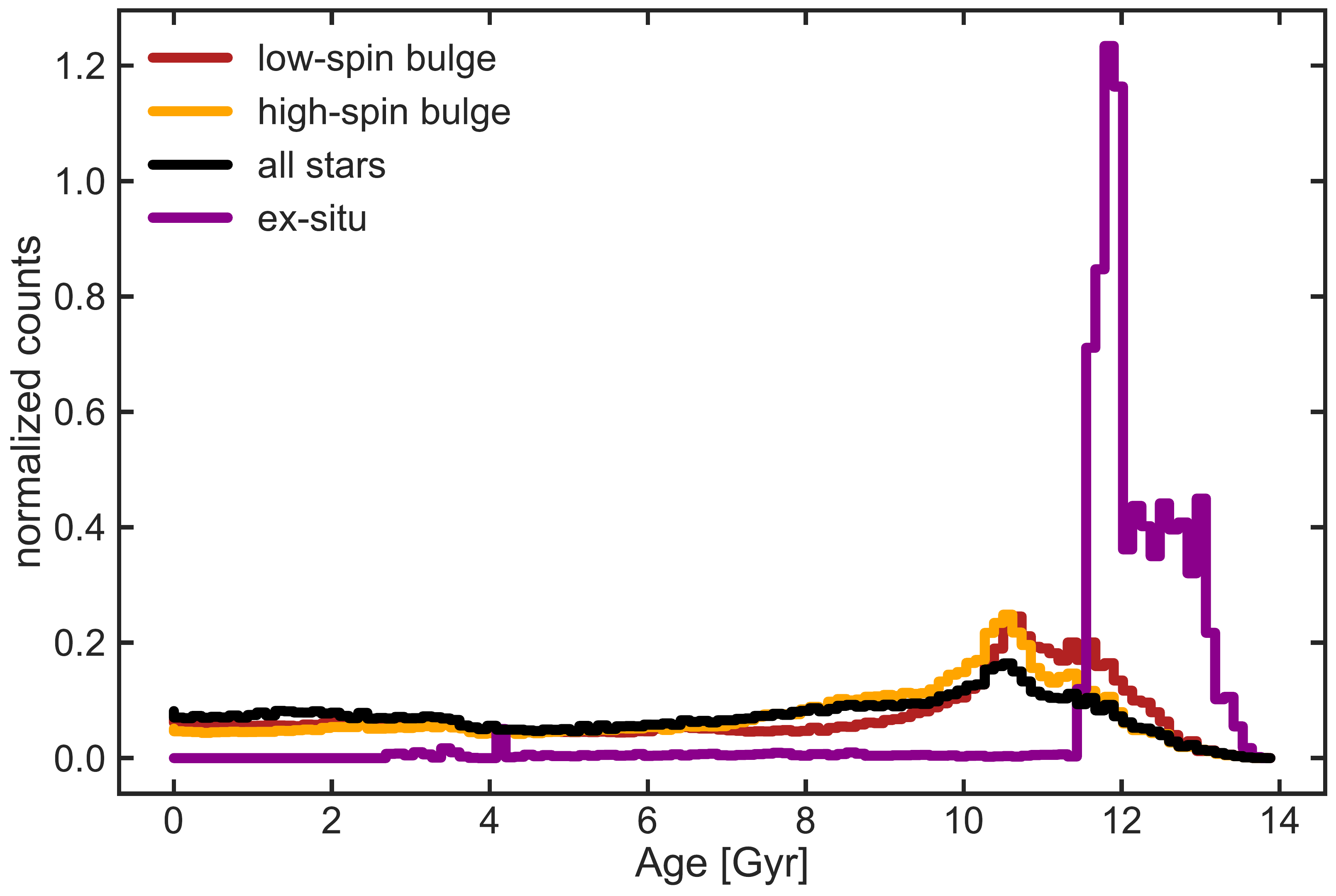}
\includegraphics[width=.5\textwidth]{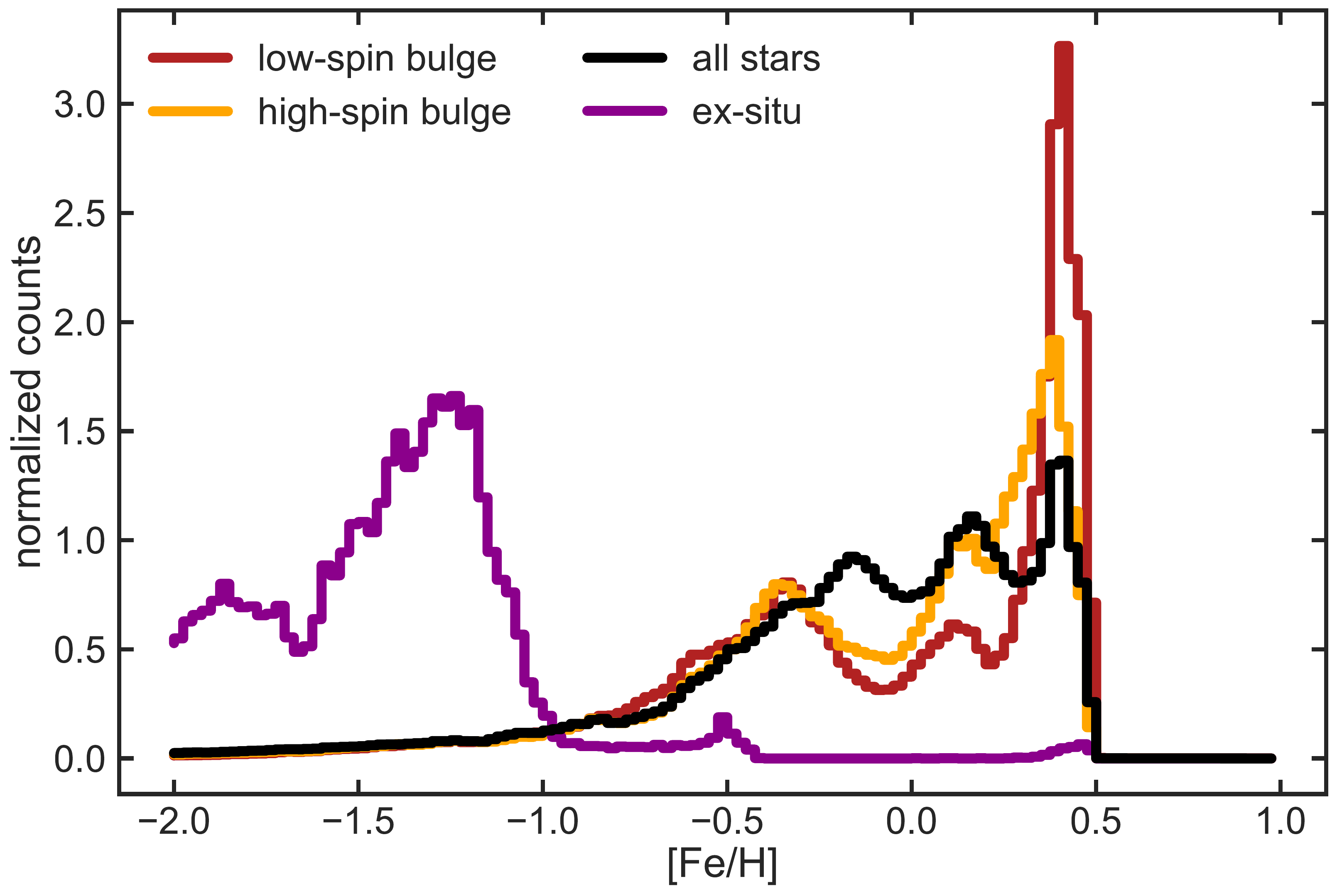}
\vspace{-.35cm}
\caption{Comparison between the age (left) and the metallicity (right) distributions of all stellar particles in the galaxy (black histogram), the two bulge components (red, orange histograms) and the ex-situ stellar particles in the two bulge components (magenta histogram). The ex-situ stars are purely old and metal-poor.}
\label{fig:ex-situ}
\end{figure*}
%%%%%%%%%%%%%%%%%%%%%%%%%%%%%%%%%%%%%%%%%%%%%%%%%%%%

% birth angular momentum
\subsubsection{Birth angular momentum}
In Fig. \ref{fig:birth_ang_mom_z} we show the total angular momentum in $z$-direction, $L_{z}$, at birth (thick lines) and at present day (thin lines), where the $z$-direction is defined as the direction of total angular momentum of the disk stellar particles. Again, the red histogram shows the low-spin bulge and the orange histogram shows the high-spin bulge component. As was already appreciated from Fig. \ref{fig:rot_disp}, the high-spin bulge shows strong rotation while the low-spin bulge is dispersion dominated. Thus, in this figure we see that the high-spin bulge has higher angular momentum at present-day compared to the low-spin bulge. Furthermore, stellar particles belonging to the high-spin bulge are on average born with a higher angular momentum compared to stellar particles of the low-spin bulge.
We also compared the total angular momentum distribution and found that the $z$-component for the high-spin bulge population is much more similar to the total angular momentum, while the low-spin bulge component has already a significant angular momentum component perpendicular to the disk at birth. This indicates that indeed the stars that end up in the high-spin bulge were part of the highly rotating thin disk component. On the other hand, the stars evolving into the low-spin bulge were part of a thicker component, with relatively larger excursions perpendicular to the disk.

% general conclusion from this findings
In summary, we find that stellar particles of both the low-spin and the high-spin bulge show very similar age and birth radii distributions because stars of both components were originally part of a disk structure. However, their birth angular momenta differ. We find that on average the angular momentum of stellar particles in the high-spin bulge is higher than for the low-spin bulge. This difference can partly be explained by the difference in radius. However, we have further compared the difference in the total angular momentum of the two bulge components. We find that the high-spin bulge stars originate from cold disk orbits,  where the angular momentum in the $z$-direction is more similar to the total angular momentum. The stars of the low-spin bulge originate from hot disk orbits, with an initial higher angular momentum component perpendicular to the disk. Other properties like the age distribution, birth positions or heights (see Fig. \ref{fig:birth_height} in the Appendix) are fairly equal for the two components. 
This suggests that internal evolution e.g. via migration in the presence of the bar, leads to a different separation of the disk populations as suggested previously by \citet[e.g.][]{Debattista2016,Fragkoudi2017}. For this particular simulation a strong bar formed about 8 to 9 Gyr ago (see paper I). 
Stellar particles initially born with lower angular momentum in the $z$-direction (higher angular momentum perpendicular to the disk) and thus initially part of a thick disk component get scattered inwards and out of the disk to evolve into the low-spin bulge. Stellar particles initially born with higher angular momentum and thus initially part of a thin disk component get more efficiently trapped in the bar and will become part of the high-spin bulge. We argue here that the bar itself is able to separate the stars into different orbit families as suggested by \cite{Debattista2016}. Especially for the younger and more metal-rich components, our analysis reveals a common disk origin of the bulge stars for both bulge components in agreement with recent studies \citep[e.g.][]{DiMatteo2016,Portail2017,Fragkoudi2018}. Thus, although kinematically distinct at present day, we find a common origin of stellar particles in the two bulge components - from the disk material. We attribute the reason for a splitting into different orbit families at the present day to a difference in birth kinematics - kinematically hot and cold disks respond differently to the formation of the bar  \citep[see also][]{Debattista2016,Fragkoudi2017}. This conclusion is further supported by the kinematics of stellar particles in the two bulge components shortly before bar formation $\sim8.7$ Gyr ago which we show in figure \ref{fig:kin}. This figure clearly shows that stellar particles in the high-spin bulge had on average higher velocities (left panel) and smaller velocity dispersions (middle and right panel) compared to the low-spin bulge prior to bar formation. Especially in the radius range $\sim2-8$ kpc where most of the stellar particles eventually joining the bar are originating from (compare figure \ref{fig:birth_pos}) is the velocity dispersion of the low-spin bulge much larger compared to the high-spin bulge.

\subsection{Origin of the ex-situ stars in the bulge}
\label{sec:merger}

%%%%%%%%%%%%%%%%%% FIGURE 14 %%%%%%%%%%%%%%%%%%%%%%%%%%%
\begin{figure*}
\includegraphics[width=.33\textwidth]{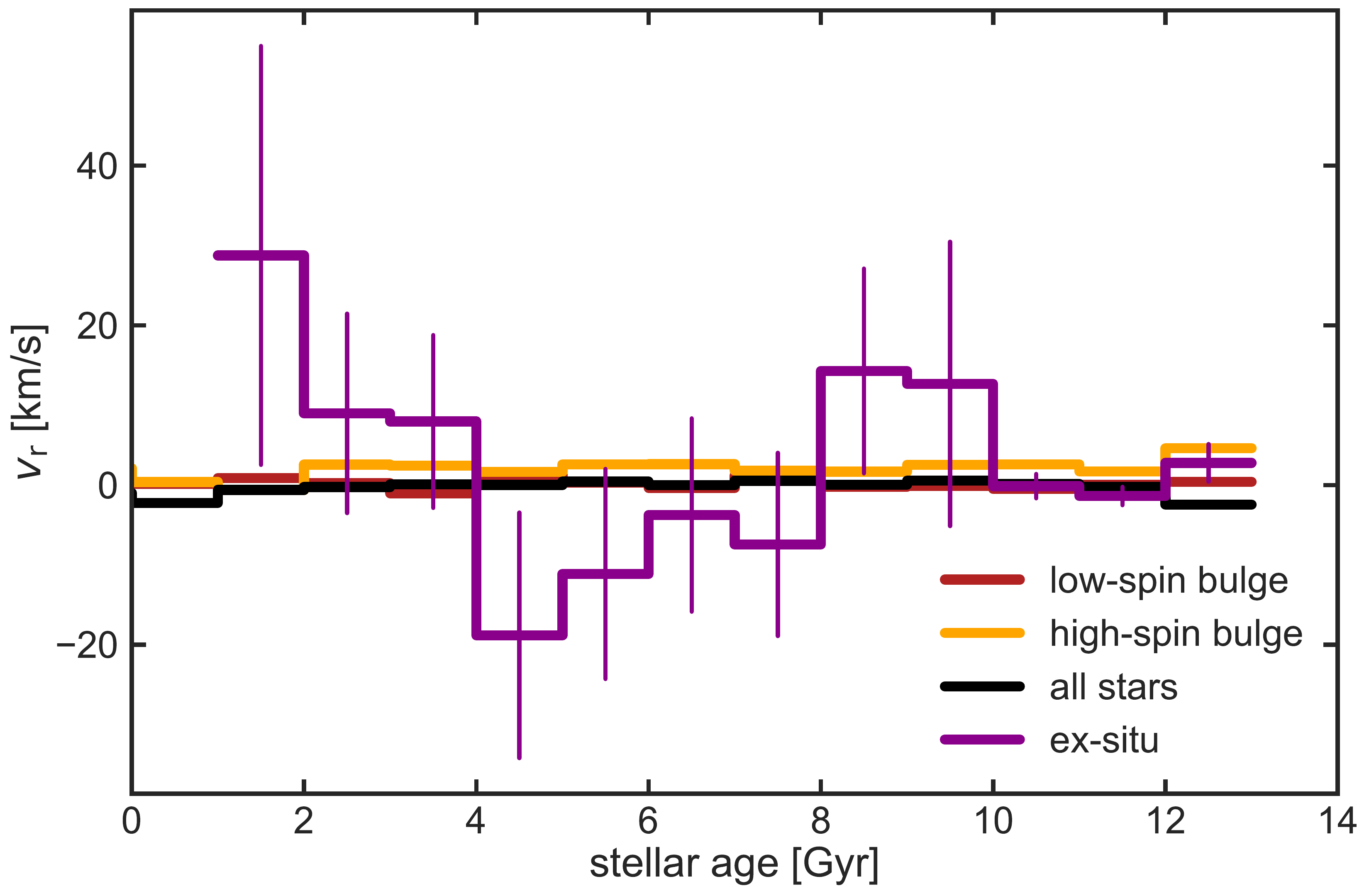}
\includegraphics[width=.33\textwidth]{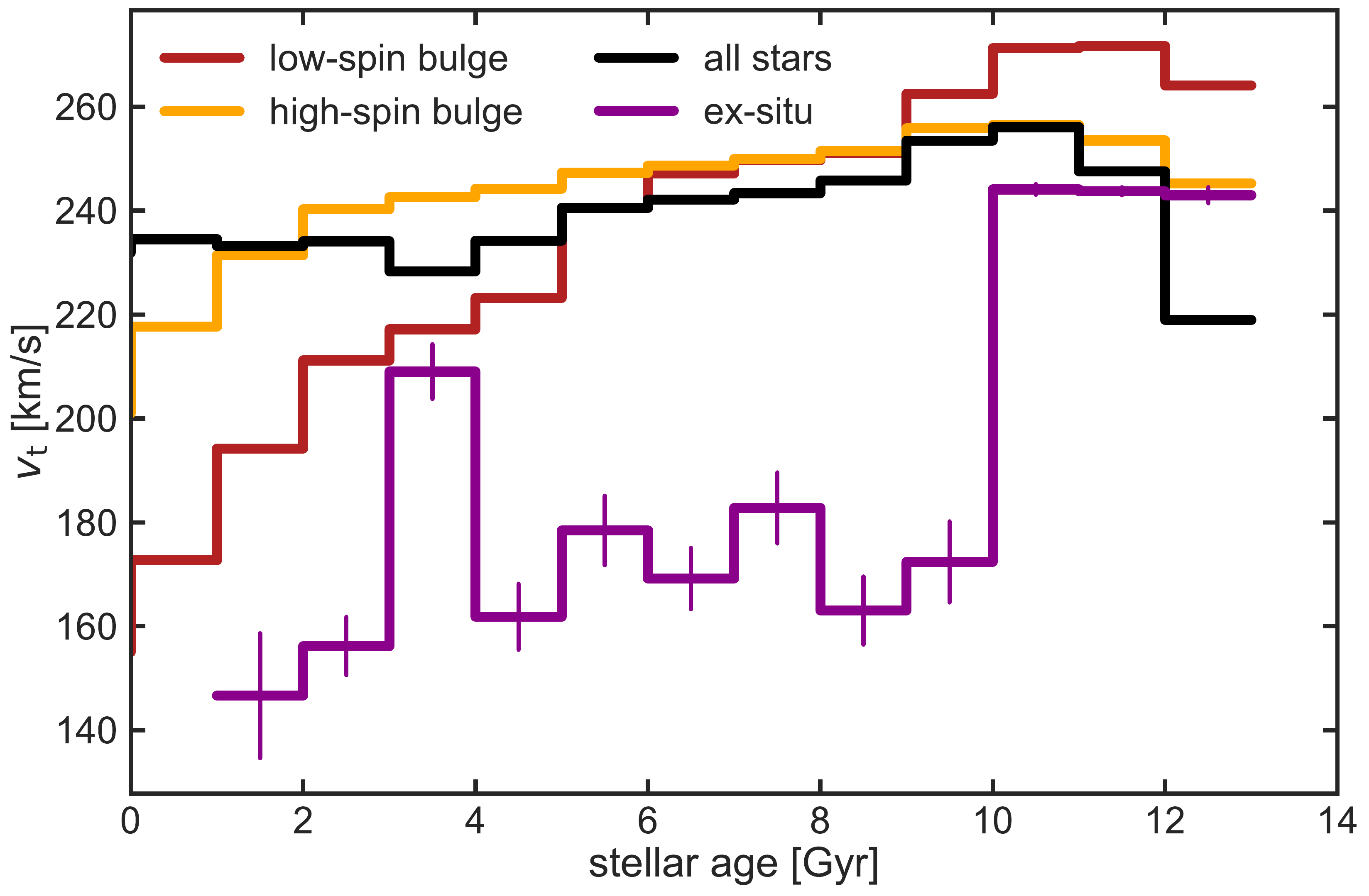}
\includegraphics[width=.33\textwidth]{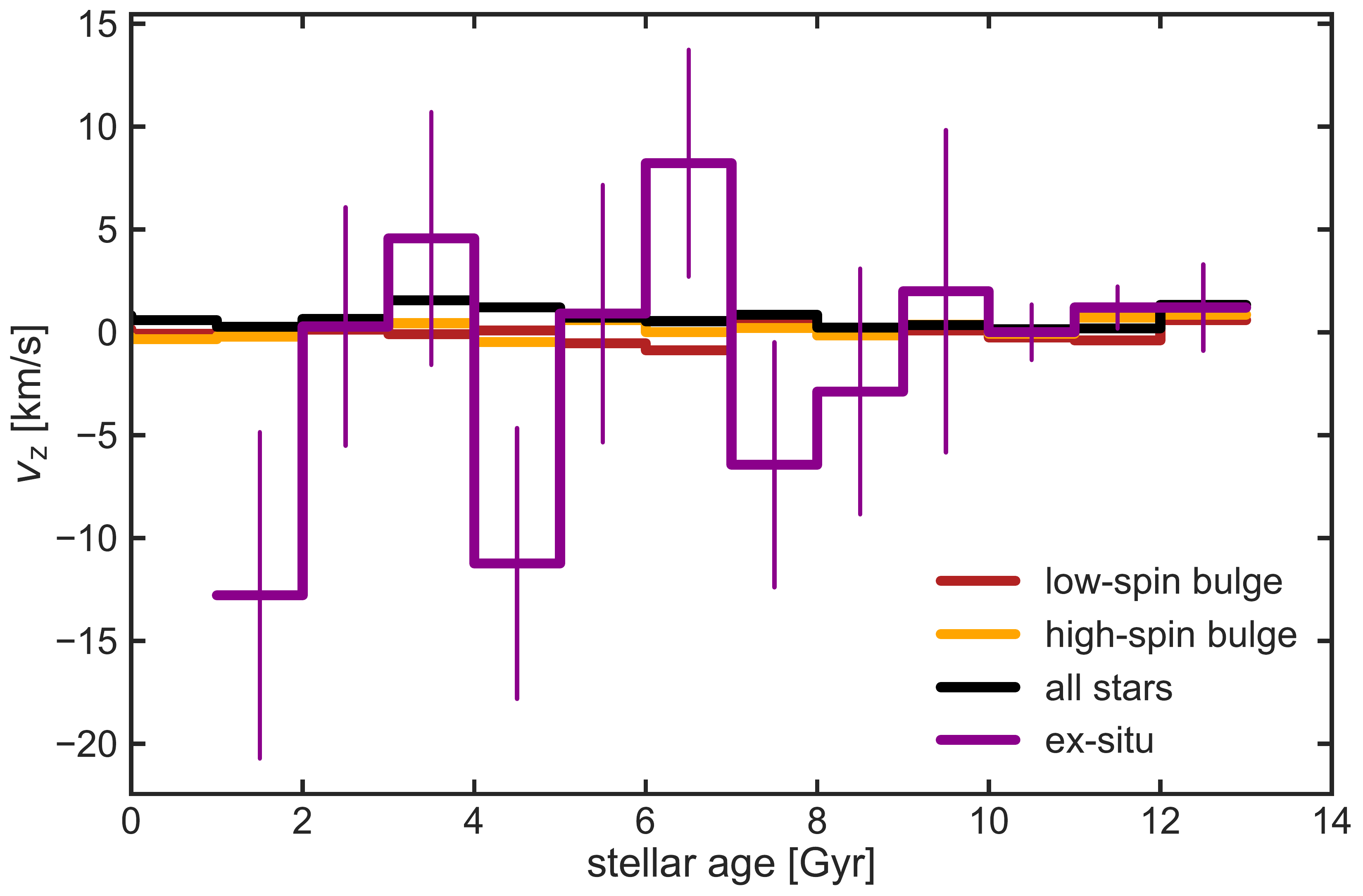}
\vspace{-.35cm}
\caption{Comparison between the mean helio-centric radial velocity (left), the mean helio-centric tangential (middle) and mean helio-centric vertical velocity (right) as a function of stellar age for all stellar particles in the galaxy (black histogram), the two bulge components (red, orange histograms) and the ex-situ stellar particles in the two bulge components (magenta histogram). The vertical purple lines indicate the uncertainty on the mean velocity for the ex-situ stars. For the other components the uncertainty is smaller than the line width.}
\label{fig:ex-situ_kin}
\end{figure*}
%%%%%%%%%%%%%%%%%%%%%%%%%%%%%%%%%%%%%%%%%%%%%%%%%%%%

The ex-situ stellar particles (born at distances $>100$ kpc from the galaxy center) in the two bulge components make up about $0.8\%$ of the low-spin bulge and about $2.3\%$ of the high-spin bulge and contribute a stellar mass of $\sim 0.7\%$ to the whole galaxy's stellar mass. Observationally, a contribution of a classical bulge component (created during the merger dominated early phase of the formation of the Galaxy) to the MW's bulge is constrained to be $<$ $8\%$ in stellar mass \citep{Shen2010Bulge}. With our simulation, we test the observational prospects to identify these ex-situ stars in the MW, if present. The ex-situ stars are preferentially found in the outskirts of the two bulge components with average galactocentric radii of about $3$ kpc (low-spin bulge) up to $8$ kpc (high-spin bulge). Thus, these stars are among the highest angular momentum stars in the two bulge components today. The ex-situ stars can not be attributed to a single accretion event but originate from several accreted and disrupted satellites over a time span of $\sim5$ Gyr previous to redshift $z\sim0.75$. A recent study by \citet{Badry2018} looked at the spatial distribution of the oldest stars in Milky Way like galaxies and found them to preferentially reside in halo like objects although their contribution in the very center of the galaxies can be significant, in very good agreement with our findings here. 

In Fig. \ref{fig:ex-situ} we show the age and metallicity distributions of the whole galaxy (black histogram), the two bulge components (red - low-spin bulge, orange - high-spin bulge) in comparison to the distributions for the ex-situ stars (magenta histogram). This figure clearly shows that the ex-situ stars are exclusively old and metal poor. A recent study by \citet{Tissera2018} looked at the hydrodynamical versions of the Aquarius simulations \citep{Scannapieco2009} to study the various properties of ex-situ and in-situ stars in the bulge region. These authors find contributions of ex-situ stars of up to $\sim40\%$ much higher than the results found here. We attribute this to different cosmic accretion histories where the galaxy analysed in this work has a very quiet merger history. Similar to our results these authors find that ex-situ stars are more metal poor compared to in-situ stars and originate from a few disrupted satellites. On the other hand, contrary to our results these authors find that ex-situ stars are slightly younger than in-situ stars.

We further tested if these ex-situ stars are identifiable through their kinematics. In Figure \ref{fig:ex-situ_kin} we show the mean helio-centric radial velocity (left panel), the mean helio-centric tangential velocity (middle panel) and mean helio-centric vertical velocity (right panel) as a function of stellar age for all the stars in the galaxy, the two bulge components and the ex-situ stars. While the radial and vertical velocities of the stars in the bulge components and the whole stellar disk average to roughly zero mean velocity in every age bin, we see that the accreted stars show on average high radial velocities of up to $30$ km/s and high vertical velocities of up to $10$ km/s. However, given the uncertainty on the radial velocity the differences between ex-situ stars and the stellar disk average is marginally significant. On the other hand, the tangential velocities in the middle panel show interesting features. The high-spin bulge rotates almost as fast as the whole disk with tangential velocities of about $230$ km/s. The low-spin bulge shows a gradient in tangential velocity across stellar age, with the oldest stars rotating more rapidly compared to the youngest stars. The ex-situ stars on the other hand show very low mean tangential velocity of (only) $160$ km/s with only the oldest stars rapidly rotating. Thus, we conclude that old metal poor stars on highly radial orbits with large vertical excursions in the MW are more likely to be of ex-situ origin in agreement with recent results derived from Gaia and SDSS data \citep{Belokurov2018}.

\section{Discussion}

We have analysed a simulation of a MW like galaxy that shows an overall boxy/peanut shaped morphology well in agreement with what is observed for the MW as we have established in paper I. Here we focused on the different stellar populations present in the inner regions of the simulation where we separated the stellar particles of the galaxy into different orbit families. This decomposition enables us to study and understand the different populations in our simulation separately. This particular galaxy has an equal mass fraction of stellar particles on spheroidal and boxy orbits, as it is shown in Figure \ref{fig:decomp_surf}. Similar results supporting our conclusions were independently found by \citep{Spinoso2017}.

Since it is widely assumed that different morphological/kinematical components of galaxies, like e.g. the high-spin or the low-spin bulge, follow from different formation scenarios \citep[e.g.][]{Kormendy2013}, we specifically focused on the origin of stars in the two bulge components. Note, we do not dispute that large low-spin bulges in galaxies are a signature of mergers, we simply demonstrate that the bar instability is able to significantly increase the vertical excursions of stars in the bulge that are in fact coming originally from the inner disk. While so far only peanut shaped bulges are thought to be secular evolution driven, low-spin bulges are associated with merger-built spheroids \citep{Binney1998,Elmegreen1998,Kormendy2013,Sellwood2014}. In this study we find that the two kinematically defined bulge components are mostly formed from disk material owing to a common secular origin as shown in Fig. \ref{fig:birth_pos}. We found that the inner disk can evolve into kinematically distinct populations of stars under the secular influence of the bar. Studying the cause of the separation of the two bulge components at present-day, we find that it is the initial angular momentum of stellar particles which influences their present-day orbits. As we have shown in Fig. \ref{fig:birth_ang_mom_z}, the birth angular momentum for the high-spin bulge is larger compared to the low-spin bulge.
Thus, we find that, except for their different orbital structure, the two bulge components share very similar properties (see Fig. \ref{fig:decomp_props}) like e.g. their metallicity distribution or age distribution. Therefore, we propose that a kinematic decomposition technique offers a fresh and distinct framework to understand and quantify the MW's bulge in times where the Gaia satellite will delivering 5+1d phase space information for millions of MW stars. This will finally enable us to apply the same techniques to both observed stars and simulated stellar particles and thus, by combining simulations and observations, to understand the origins of the Bulge.

%%%%%%%%%%%%%%%%% TABLE 1 %%%%%%%%%%%%%%%%%%%%%%%%%%%
\begin{table}
\label{tab:pop}
\begin{center}
\caption{Gaussian component fit to metallicity distribution.}
\begin{tabular}{c c c c}
		\hline\hline
		population & mean [Fe/H] & weight & $\sigma$ \\
		\hline
        & low-spin bulge & & \\
        \hline
		1 & 0.43 & 0.35 & 0.05\\
        2 & 0.16 & 0.20 & 0.13\\
        3 & -0.32 & 0.27 & 0.15\\
        4 & -0.68 & 0.12 & 0.21\\
        5 & -1.31 & 0.06 &  0.44\\
        \hline
        & high-spin bulge & & \\
        \hline
        1 & 0.38 & 0.30 & 0.07\\
        2 & 0.13 & 0.26 & 0.12\\
        3 & -0.31 & 0.30 & 0.16\\
        4 & -0.79 & 0.09 & 0.25\\
        5 & -1.46 & 0.05 & 0.46\\
		\hline
\end{tabular}
\end{center}
\end{table}
%%%%%%%%%%%%%%%%%%%%%%%%%%%%%%%%%%%%%%%%%%%%%%%%%%%%%

Already with the existing data sets, like e.g. the ARGOS survey, we are able to compare the different kinematic signatures of the two bulge components to observations of the MW. Comparing individual and combined rotation and dispersion profiles we find that only the combination of the two bulge components matches the observed kinematics of bulge stars from the ARGOS survey (see Fig. \ref{fig:rot_disp}). This finding makes us conclude that also for the MW we might find stars on both orbit families once we have access to stellar orbits through Gaia. Furthermore, we expect to find stars on both orbits across all metallicities as we have shown that the age and metallicity distributions for both bulge populations are broad (see Fig. \ref{fig:decomp_props}). This feature is a direct result of the continuous star formation in the disk and the subsequent inclusion of further disk material into the bulge component by secular evolution. However, a broad metallicity distribution is also found for the stars included in the ARGOS survey \citep{Ness2013a}. In order to understand similarities and differences between the ARGOS survey and our simulation we briefly discuss the metallicity distributions in slightly more detail below.

\subsection{Metallicity distribution of the ``low-spin bulge" and the ``high-spin bulge"}
\label{sec:met}

%%%%%%%%%%%%%%%%%% FIGURE 15 %%%%%%%%%%%%%%%%%%%%%%%%%%%

\begin{figure}
\includegraphics[width=\columnwidth]{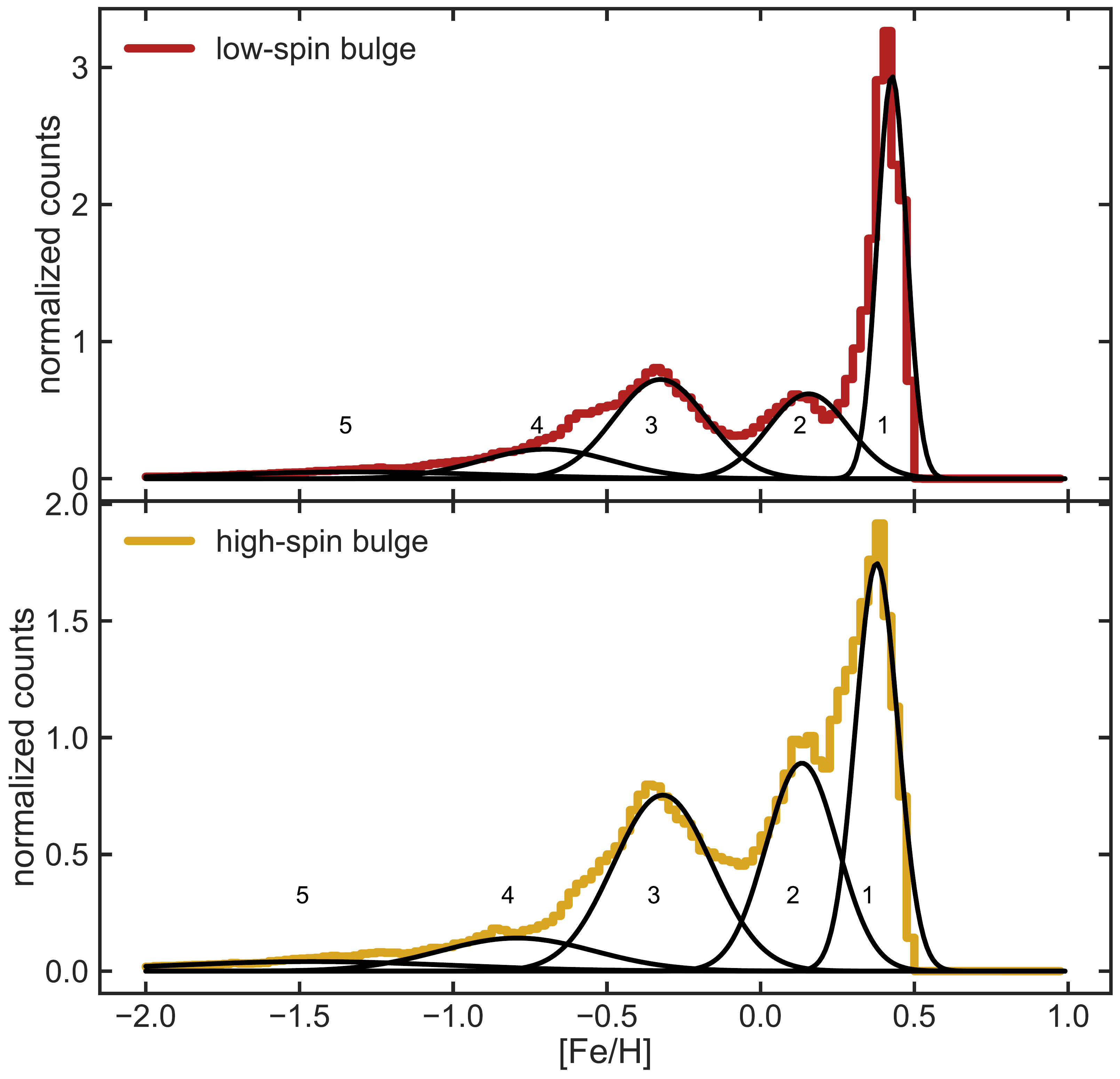}
\vspace{-.35cm}
\caption{Chemical sub-components of the kinematically decomposed low-spin bulge (top panel) and high-spin bulge (bottom panel). The metallicity of the components spans the entire Metallicity Distribution Function of the bulge. Whilst we can identify 5 different sub-components 1 to 5 using Gaussian mixture models, as indicated, we do not associate these with the discrete populations as per \cite{Ness2013a}. In Table \ref{tab:pop} we show the mean, the standard deviation and the weight of each metallicity component.  
}
\label{fig:met_pop_decomp}
\end{figure}
%%%%%%%%%%%%%%%%%%%%%%%%%%%%%%%%%%%%%%%%%%%%%%%%%%%%

In the two left most panels of Fig. \ref{fig:decomp_props} we have seen that the low-spin and the high-spin bulge populations show distinct substructures in their metallicity and age distributions. These features are similar to what is observed in the ARGOS data set \citep[see][]{Ness2013a} and observations from \citet[][]{Clarkson2008,Hill2011,Gonzalez2011}. Therefore, we examine the substructure in the metallicity distribution of our simulation following the approach by \citep{Ness2013a} using a Gaussian mixture model. Similar to the results of \citet{Ness2013a} for the ARGOS data we are able to identify 5 different sub-components (1 to 5) in our simulation. The metallicity distribution function (MDF) of both bulge components together with the decomposition (black lines) are shown in Fig \ref{fig:met_pop_decomp} and parameters for the metallicity decomposition are shown in Table \ref{tab:pop}.

The different components in the MDF of the MW bulge span each a small, overlapping part of the MDF. \citet{Ness2013a} associated the components each with different populations in the MW, including bulge, thick disk and halo stars. In contrast to that, we find that each of our bulge populations spans the entire range of the MDF which shows substructure and can itself be broken up into sub-populations. This highlights that metallicity itself in both observations and in simulations may in fact be sub-optimal for understanding the evolution history of the bulge. In simulations, due to the uncertainty in recreating the chemical enrichment history as a function of time and in observations due to stars at a given metallicity reflecting a range of different times of formation -- and not a direct tracer of history \citep{Minchev2017}. Nevertheless, since the morphology of the central regions of the MW is observed to be metallicity dependent  we briefly discuss this for our simulation as well. 

\subsection{Morphology}

%%%%%%%%%%%%%%%%%% FIGURE 16 %%%%%%%%%%%%%%%%%%%%%%%%%%%
\begin{figure*}
\includegraphics[width=\textwidth]{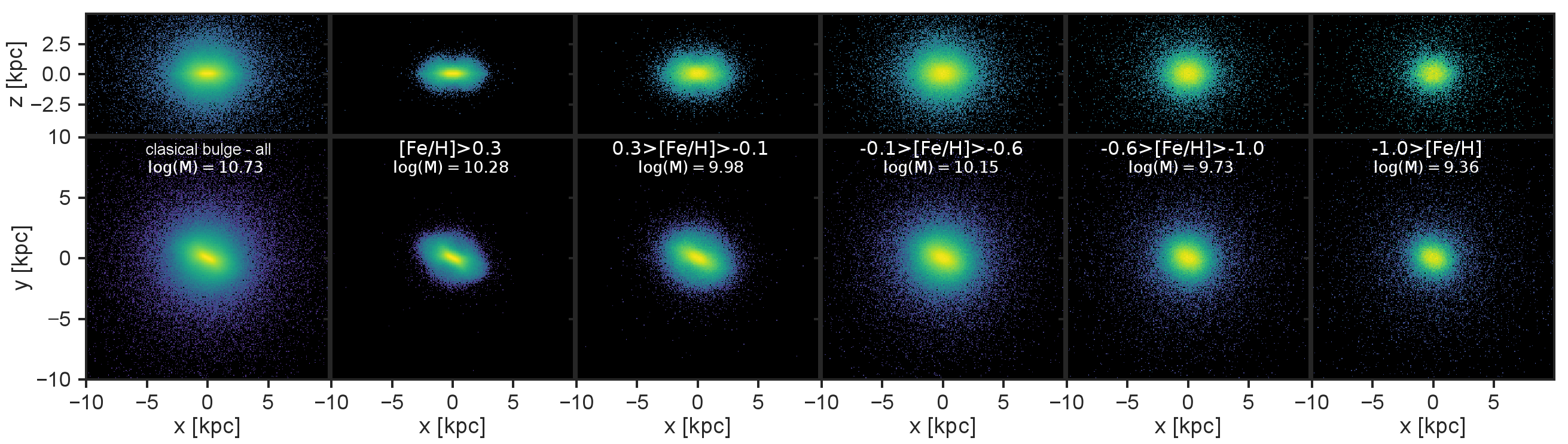}
%\vspace{-.35cm}
\caption{Surface density maps in edge-on (top row) and face-on (bottom row) projections for all stars (left most column) in the low-spin bulge component and for stars in different metallicity bins (second column from the left to most right column) as indicated in the panels. Each panel indicates the amount of stellar mass in solar masses contained in the component.
}
\label{fig:cb_surf_den}
\end{figure*}
%%%%%%%%%%%%%%%%%%%%%%%%%%%%%%%%%%%%%%%%%%%%%%%%%%%%

%%%%%%%%%%%%%%%%%% FIGURE 17 %%%%%%%%%%%%%%%%%%%%%%%%%%%
\begin{figure*}
\includegraphics[width=\textwidth]{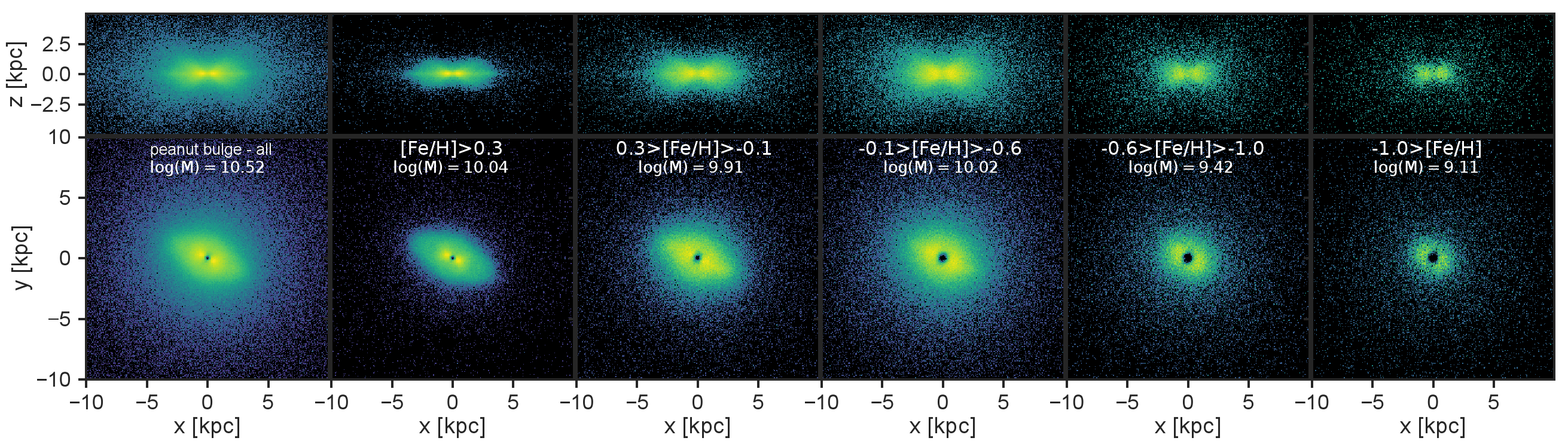}
%\vspace{-.35cm}
\caption{Same as Fig. \ref{fig:cb_surf_den} but for the high-spin bulge component. 
}
\label{fig:pb_surf_den}
\end{figure*}
%%%%%%%%%%%%%%%%%%%%%%%%%%%%%%%%%%%%%%%%%%%%%%%%%%%%

The substructure in abundance space directly relates to substructure in geometrical space. Recognising the different morphologies as a function of metallicity for stellar particles all on similar orbits is important in interpreting observational results. In Fig. \ref{fig:cb_surf_den} we show the different metallicity components of the low-spin bulge and in Fig. \ref{fig:pb_surf_den} of the high-spin bulge, respectively.

Across all metallicity bins, [Fe/H] $>$ --1.0 dex to [Fe/H] $<$ 0.3 dex, the low-spin bulge (Fig. \ref{fig:cb_surf_den}) is more spherically symmetric compared to the high-spin bulge (Fig. \ref{fig:pb_surf_den}) which has a distinct boxy- or bar-like morphology\footnote{The small hole in the center of the high-spin bulge is due to our decomposition scheme where stars in the very center get assigned to the low-spin bulge component}. Furthermore, in edge-on view there is a clear X-shape morphology present across all metallicities for the high-spin bulge. We see further, that the more metal poor components of the low-spin bulge are more spherically symmetric compared to their higher metallicity counterparts in agreement with recent findings by \citet{Zoccali2018}. When looking at the kinematics of stars in the different metallicity sub-components we find that the two highest metallicity components of the low-spin bulge show some net rotation while the three lowest metallicity components are purely non-rotating. On the other hand, all metallicity components of the high-spin bulge are rotating at equal speed (compare figures \ref{fig:maps_decomp_pb}, \ref{fig:maps_decomp_cb} in the Appendix). Combining both, the low-spin and the high-spin bulge stars and investigating their combined kinematics (see fig. \ref{fig:maps_decomp_both} in the Appendix) we find that the three highest metallicity components show similar line-of-sight velocities. Given the overall offset in stellar metallicities between the simulation and the MW with the simulation shifted to higher metallicities this in qualitative agreement with the results from \citet{Ness2013b}.

The stars in the low-spin bulge are most spherically symmetric for the most metal-poor (oldest) stars and show an elongation in the xy-plane, appearing increasingly ellipsoidal, following the bar outline, at higher metallicities (the youngest stars).
In contrast, the most metal-rich (youngest) stars of the high-spin bulge show the strongest signature of a thin bar-like feature close to the mid-plane which extends out to $\sim5$ kpc. Interestingly, this is a similar extent to the long-bar proposed by \cite{Portail2015}. The thickness of the X-shape increases with decreasing metallicity, to the intermediate metallicity bin, while the metal poorest stars are again more centrally concentrated. The bar signature diminishes with decreasing metallicity. This is similar to what is found by \cite{Portail2017a} for their model, and to what is observed for the old Mira population in the MW \citep{Lopez2017}. This finding agrees further with recent results for bulges in external galaxies \citep{Seidel2015} where younger (and thus more metal rich) stars more strongly trace the bar feature.

Finally, if we compare the mass fraction of stars belonging to the two lowest metallicity, non-rotating components of the low-spin bulge (which are also the oldest ones) to what is constrained for the MW, we find good agreement. Components 4 and 5 of the low-spin bulge account together for $\sim 6\%$ of the disk mass, while for the MW the mass fraction of an old merger-generated bulge component is constrained to be less than $\sim8\%$ \citep{Shen2010Bulge}. \citet{Debattista2016} find that their model reproduces the MW only if they add $\sim10\%$ of old, non-rotating stars with [Fe/H]$<-1.0$. This is in good agreement with the mass fraction of the two lowest metallicity, non-rotating components of the low-spin bulge which amounts to $\sim8\%$ in our simulation.

\section{Summary and Conclusion}

In this paper we analysed the fully cosmological hydrodynamical simulation of a MW like galaxy which shows bulge properties that agree well with the observed properties of the MW bulge as described in \citet{Buck2018}. We employ the kinematic decomposition of \citet{Obreja2018} to separate the galaxy into its constituent (kinematically) distinct stellar populations. These include two bulge populations, two disk populations and a stellar halo. We use this kinematic decomposition to study, in detail, the different stellar components of the two bulge components, focusing on their kinematical, temporal, and chemical properties as well as their origin. We compare the kinematics of stellar particles in the bulge region of our simulation with observed kinematics of the MW bulge stars and find qualitative good agreement. Our main results can be summarised as follows:

\begin{itemize}
\item Applying the kinematical decomposition algorithm to our galaxy we find five different components: two bulges, a high-spin bulge and a low-spin bulge (see Fig. \ref{fig:decomp_surf} and Fig. \ref{fig:decomp_props}), a low- and a high-spin disk, and a stellar halo.  
\item We investigate the stellar kinematics in the two bulge components (Fig. \ref{fig:rot_disp}) and compare it to observed kinematics of MW bulge stars. We find that the low-spin bulge is mostly slow or non-rotating with a latitude dependent dispersion profile while the high-spin bulge is fast rotating and shows only a very small latitude dependence --- as is observed for the MW (Fig. \ref{fig:cb_rot_disp_map}). 
\item We find that single components can not match the observed rotation and dispersion profiles. However, the kinematics of both bulge components together well matches the observations. 
\item We find remarkable similarity for the birth position of stars in the low-spin and the high-spin bulge (Fig. \ref{fig:birth_pos}) despite their different morphologies at the present day indicating a common disk origin. However, high-spin bulge stars have higher birth angular momentum compared to the low-spin bulge component (Fig. \ref{fig:birth_ang_mom_z}) pointing towards a thin and a thick disk origin of the two components, repsectively. We conclude that it is the secular evolution under the influence of the bar which separates initially co-spatial  populations into different orbit families. 
\item We analyse the formation mechanism of the two bulge populations. We find that these populations are almost entirely formed from the disk. Thus, the assumption that a spherical morphology originates via merger driven formation is not supported by our simulation. A boxy/peanut morphology is the signature of formation from the disk but a spheroidal morphology can also be formed from the disk. This also means that a group of stars on spherical orbits in the bulge have not necessarily have formed via mergers. 
\item A small stellar population of the two bulges was formed via mergers at very early times. We can identify these stars in the simulation as we have the full galactic formation history in the cosmological context. These stars are characterized by being exclusively old and metal poor and on highly radial orbits. Thus, they are preferentially found in the outskirts of the bulge, at galactocentric radii of $\sim3$ kpc.
\item The two bulge populations show distinct substructure in their (overlapping) metallicity distributions and a metallicity dependent morphology. For both bulge components, the vertical thickness increases with decreasing metallicity, similar to what is reported in \citet{Ness2013a}. We find that stellar metallicity alone can not be used as proxy for different kinematical components. This implies that full kinematical data from future surveys is needed in order to disentangle the different populations in the MW bulge.
\end{itemize}

%%%%%%%%%%%%%%%%%%%%%%%%%%%%%%%%%%%%%%%%%%%%%%%%%%%
\acknowledgments
\section*{Acknowledgments}
%%%%%%%%%%%%%%%%%%%%%%%%%%%%%%%%%%%%%%%%%%%%%%%%%%%
We thank the anonymous referee for useful comments and suggestions which improved the quality and readability of this paper.
We like to thank Ortwin Gerhard, Paola Di Matteo and Francesca Fragkoudi for very useful discussions and suggestions on an early version of this draft. TB thanks Hans-Walter Rix for very fruitful discussions and helpful and inspiring comments on this work.
TB acknowledges support from the Sonderforschungsbereich SFB 881 “The Milky Way System” (subproject A2) of the German Research Foundation (DFG). AO is funded by the Deutsche Forschungsgemeinschaft (DFG, German Research Foundation) -- MO 2979/1-1. The authors gratefully acknowledge the Gauss Centre for Supercomputing e.V. (www.gauss-centre.eu) for funding this project by providing computing time on the GCS Supercomputer SuperMUC at Leibniz Supercomputing Centre (www.lrz.de). This research was carried out on the High Performance Computing resources at New York University Abu Dhabi; Simulations have been performed on the ISAAC cluster of the Max-Planck-Institut fuer Astronomie at the Rechenzentrum in Garching and the HYDRA and DRACO cluster at the Rechenzentrum in Garching. We greatly appreciate the contributions of all these computing allocations. 
This research made further use of the {\sc{pynbody}} package \citet{pynbody} to analyze the simulations and used the {\sc{python}} package {\sc{matplotlib}} \citep{matplotlib} to display all figures in this work. Data analysis for this work made intensive use of the {\sc{python}}  library  {\sc{SciPy}} \citep{scipy}, in particular {\sc{NumPy and IPython}} \citep{numpy,ipython}.
 
\bibliography{astro-ph.bib}

\appendix
\label{appendix}
%\renewcommand{\thefigure}{A\arabic{figure}}
%\setcounter{figure}{0}
%\vspace*{-.5cm}
%%%%%%%%%%%%%%%%%%%%%%%%%%%%%%%%%%%%%%%%%%%%%%%%%%%
%\section*{Appendix:}
%%%%%%%%%%%%%%%%%%%%%%%%%%%%%%%%%%%%%%%%%%%%%%%%%%%

\subsection{Selection of the number of components}
\begin{figure}
\centering
\includegraphics[width=.3\textwidth]{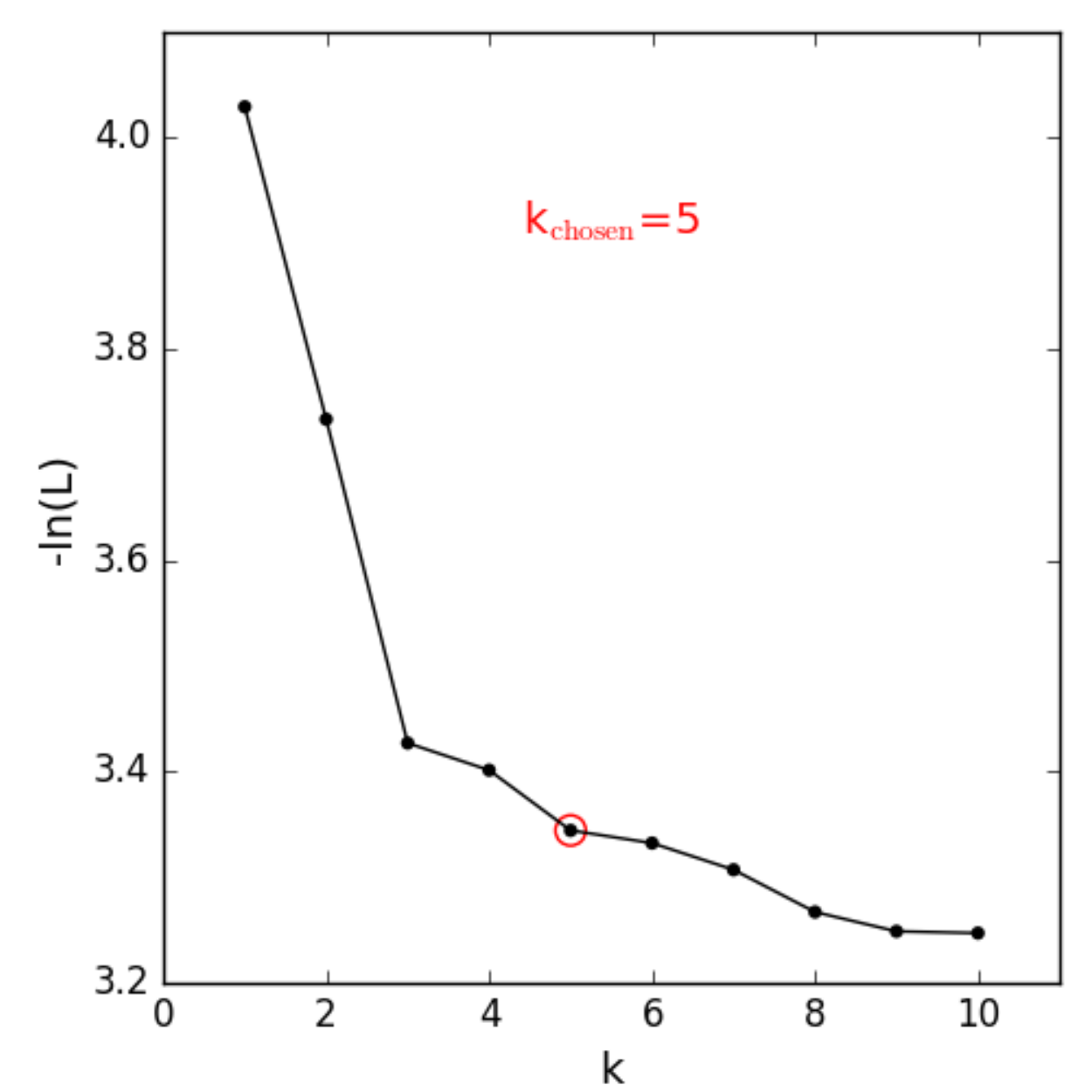}
\vspace{-.35cm}
\caption{Negative log likelihood $-\log\left(L\right)$ as a function of the number of components $k$.}
\label{fig:bic}
\end{figure}

The choice of five components is essentially based on the visual inspection of the maps of edge-on surface mass density and line-of-sight velocity. While we have also studied if various information criteria like the BIC \citep{Schwarz1978} or AIC \citep{Akaike1974} can be used for model selection in this context we found that none of these criteria turned out to be suitable for our problem. Both the BIC and AIC decrease with the number of components, however their minima are shallow and at large values of $k$. Given the very large data sample ($\sim10^6$ particles) and the fact that we use GMM in a 3D space, it is not unexpected that these criteria are not useful for our problem. Fig. \ref{fig:bic} shows how the negative log likelihood ($-\log\left(L\right)$) varies as a function of $k$. The BIC and the AIC (not shown) are just scaled version of the $-\log\left(L\right)$ because their penalty functions are subdominant over the log likelihood term. The problem of finding an appropriate information criteria is also complicated by the fact that the various components can have an important amount of overlap in this 3D space.

%%%%%%%%%%%%%%%%%% FIGURE A2 %%%%%%%%%%%%%%%%%%%%%%%%%%%
%%%%%%%%%%%%%%%%%%%%%%%%%%%%%%%%%%%%%%%%%%%%%%%%%%%%

\subsection{Distribution of dynamical features}
\begin{figure*}
\includegraphics[width=.33\textwidth]{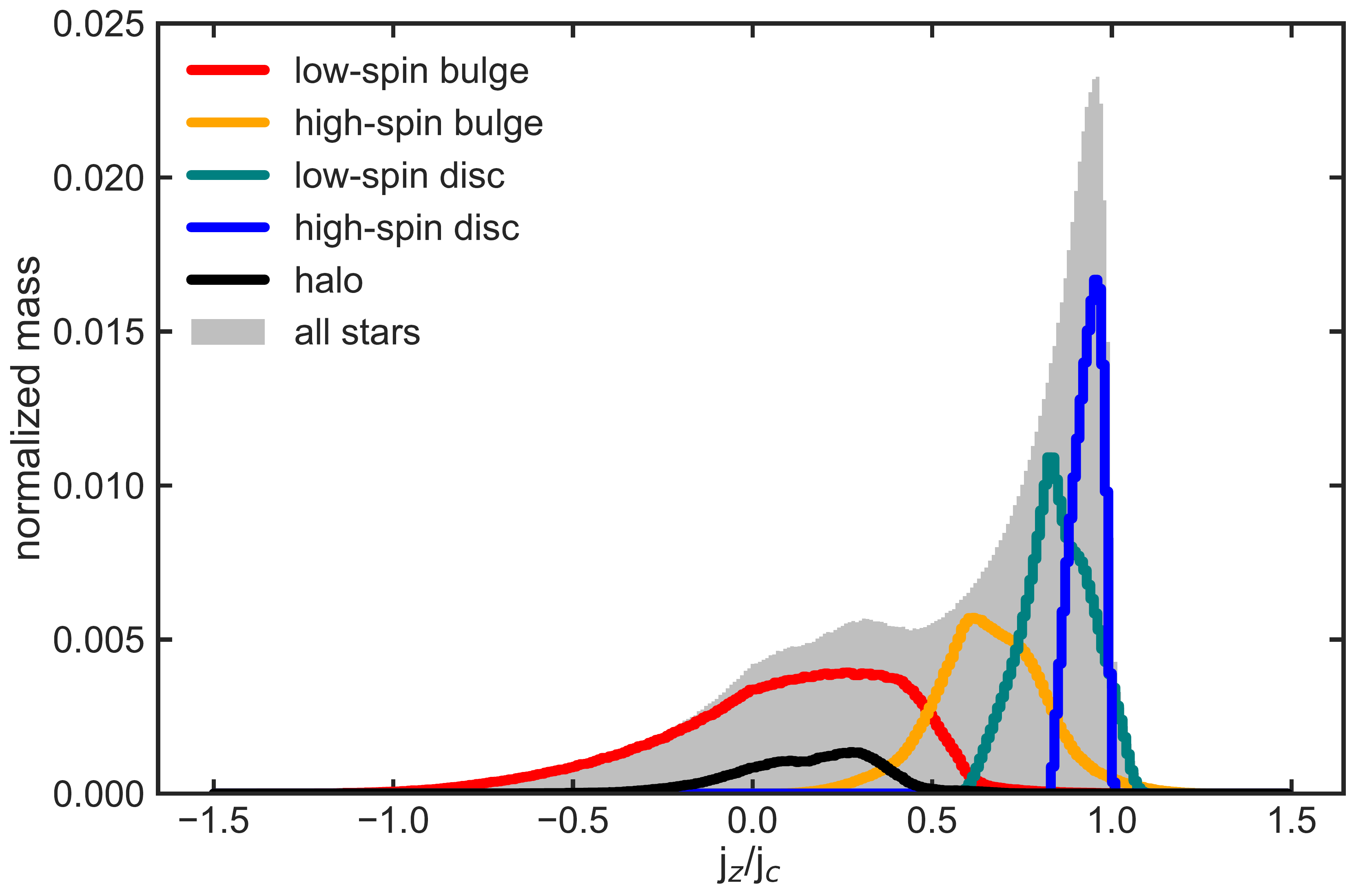}
\includegraphics[width=.33\textwidth]{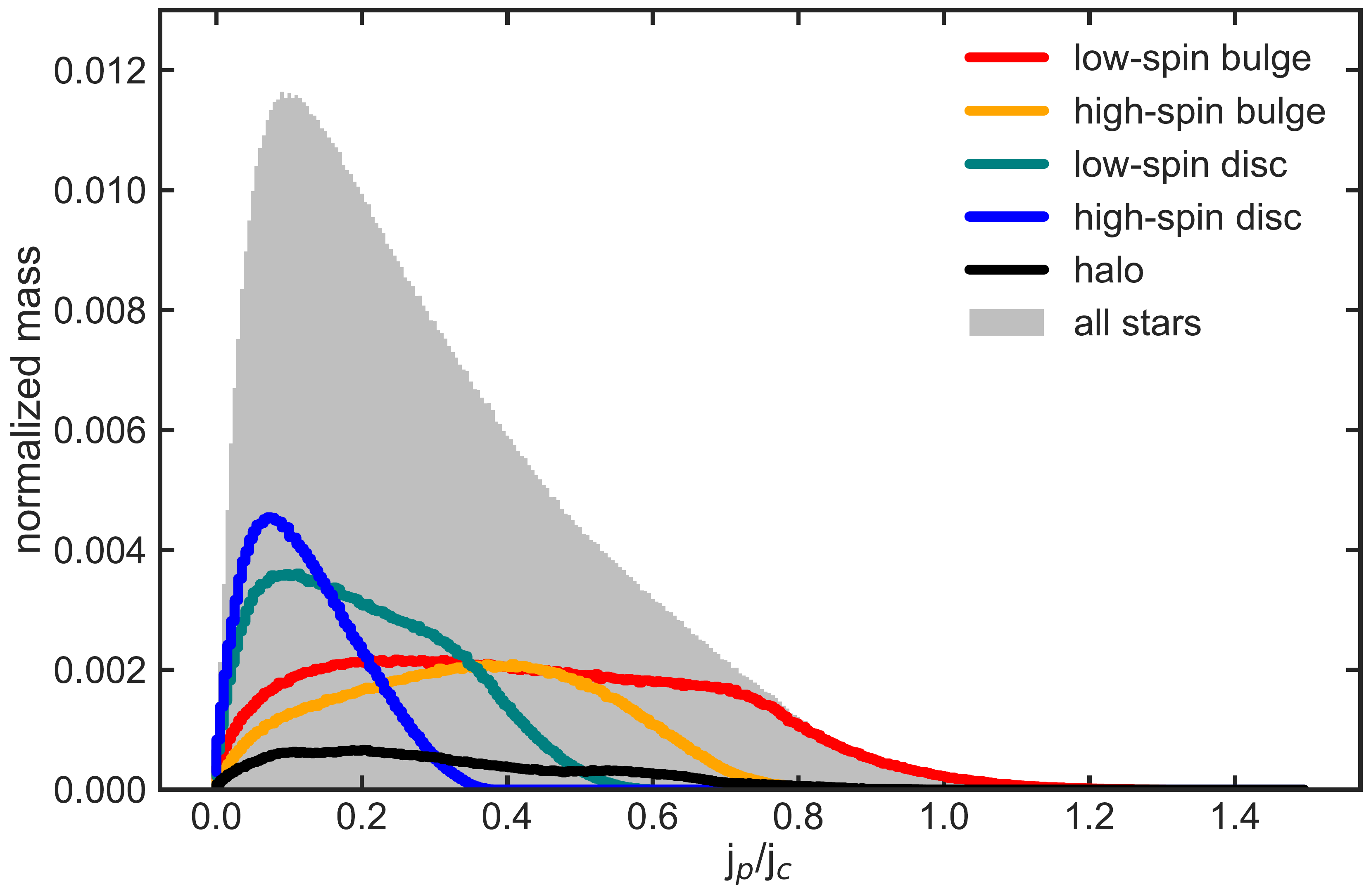}
\includegraphics[width=.33\textwidth]{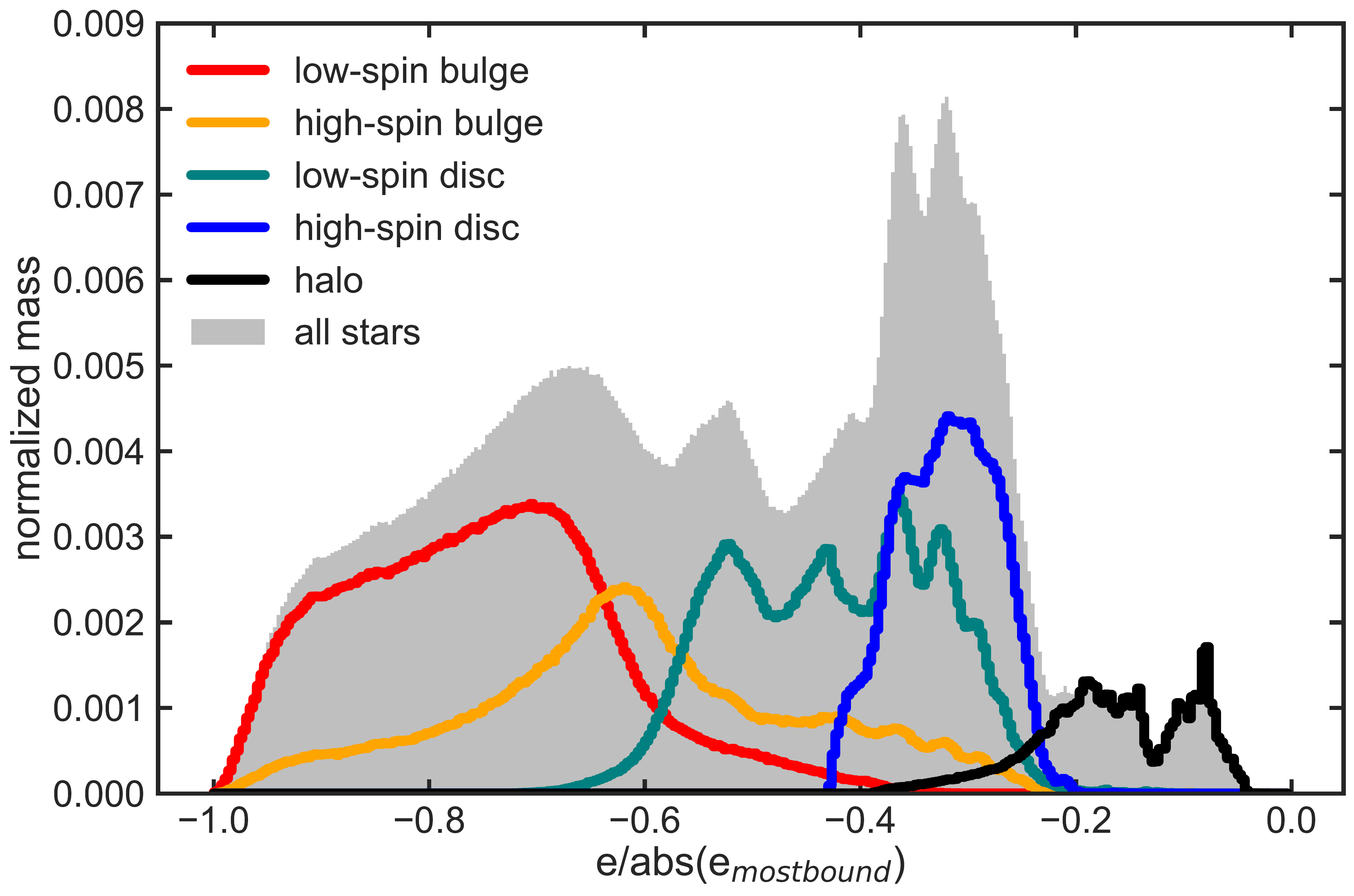}
%\vspace{-.25cm}
\caption{The result of the GMM decomposition shown as normalized stellar mass in each component (colored histograms) as function of dynamical features: $j_z/j_c$ (left panel), $j_p/j_c$ (middle panel), $e/\vert e_{\rm{mostbound}} \vert$ (right panel). The gray histograms show the normalized stellar mass distribution for the whole stellar population of the galaxy.
}
\label{fig:decomp_distr}
\end{figure*}

Figure \ref{fig:decomp_distr} presents the distribution of $j_z/j_c$ (left panel), $j_p/j_c$ (middle panel), $e/\vert e_{\rm{mostbound}} \vert$ (right panel) for the whole stellar population in the galaxy as well as for the five sub-components after applying the decomposition algorithm.
%%%%%%%%%%%%%%%%%% FIGURE A1 %%%%%%%%%%%%%%%%%%%%%%%%%%%

%%%%%%%%%%%%%%%%%%%%%%%%%%%%%%%%%%%%%%%%%%%%%%%%%%%%

\subsection{Kinematics of decomposed components}

\begin{figure*}
\includegraphics[width=\textwidth]{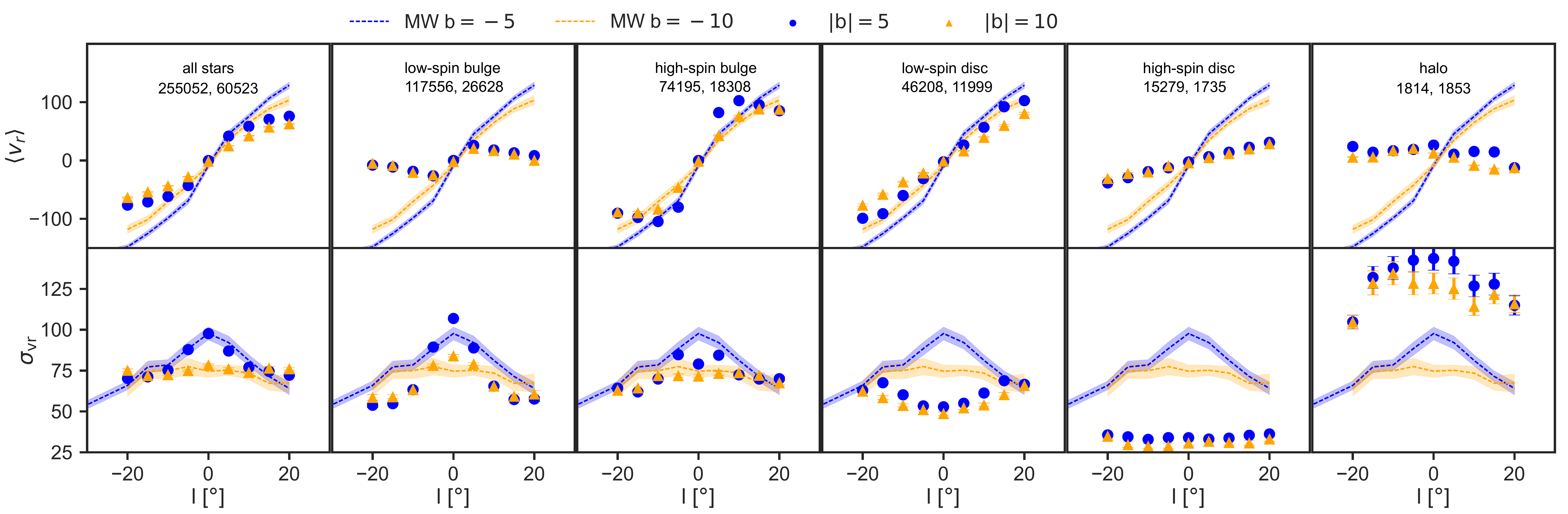}
\vspace{-.35cm}
\caption{Same as Fig. \ref{fig:rot_disp} but this time different columns show the different components as indicated with labels in the panels of the first row instead of different age populations. 
}
\label{fig:rot_disp_decomp}
\end{figure*}

The dispersion profiles of low-spin bulge and high-spin bulge are distinct and clearly separate from the dispersion profile of the halo component (compare Fig. \ref{fig:rot_disp_decomp}). Our halo component has very similar dispersion to the RR Lyrae population int he inner region \cite{Kunder2016}.

%%%%%%%%%%%%%%%%%% FIGURE A2 %%%%%%%%%%%%%%%%%%%%%%%%%%%
%%%%%%%%%%%%%%%%%%%%%%%%%%%%%%%%%%%%%%%%%%%%%%%%%%%%

%%%%%%%%%%%%%%%%%% FIGURE A3 %%%%%%%%%%%%%%%%%%%%%%%%%%%
\begin{figure*}
\centering
\includegraphics[width=.75\textwidth]{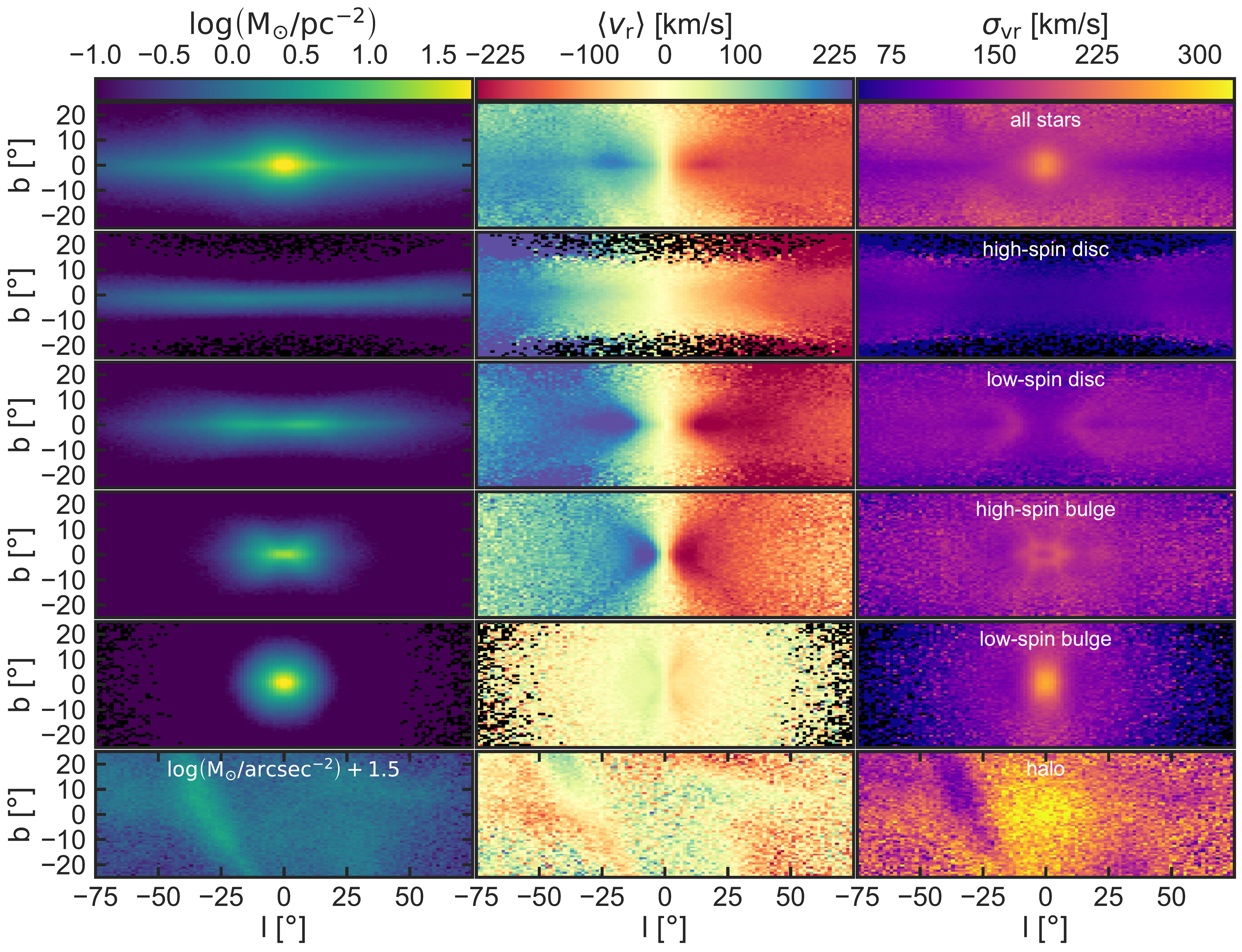}
\vspace{-.35cm}
\caption{Surface density (left column), mass-weighted rotation (middle column) and dispersion maps (right column) for different kinematically defined stellar  populations in $(l,b)$-projection for stars within $R=25$ kpc from the center of our simulated galaxy.  The \emph{first row} shows the results for all stars, the \emph{second row} shows low-spin disk stars, the \emph{third row} shows low-spin disk stars, the \emph{fourth row} shows stars in the high-spin bulge and the \emph{bottom row} shows stars of the low-spin bulge.}
\label{fig:maps_decomp}
\end{figure*}
%%%%%%%%%%%%%%%%%%%%%%%%%%%%%%%%%%%%%%%%%%%%%%%%%%%%

\subsection{Birth height of bulge stars}
\begin{figure}
\centering
\includegraphics[width=.5\textwidth]{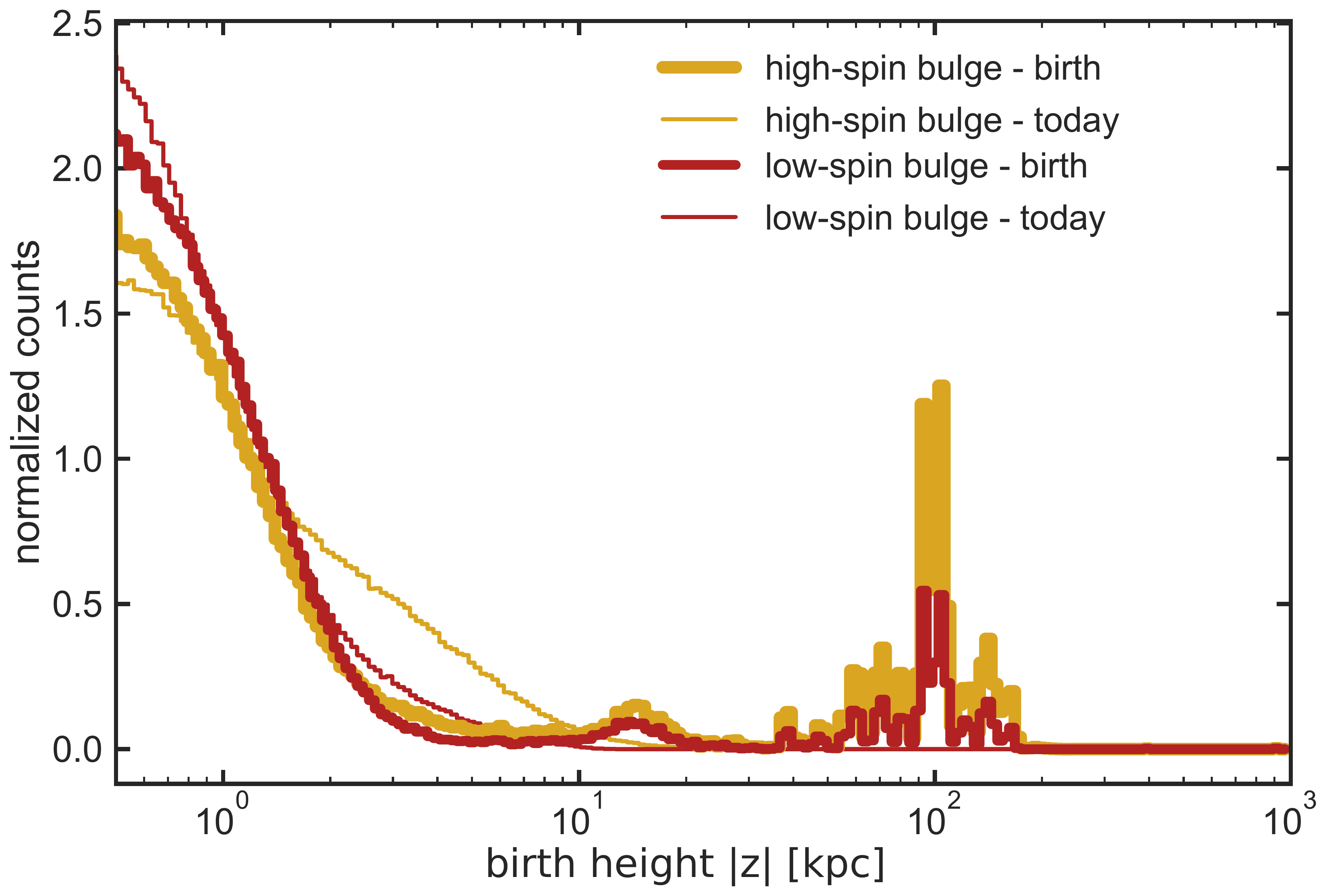}
%\vspace{-.35cm}
\caption{Comparison between the birth height from the disk mid-plane $\vert z\vert$ (thick lines) and the present day height (thin lines) above the stellar disk (in the $x-y$-plane) for stellar particles in the low-spin bulge (red histogram) component and the high-spin bulge component (orange histogram).}
\label{fig:birth_height}
\end{figure}

Figure \ref{fig:birth_height} compares the height $\vert z\vert$ above the stellar disk (in the x-y-plane) of stellar particles in the high-spin bulge and in the low-spin bulge at birth time and at present day. Both components show very similar distributions in their birth heights.
%%%%%%%%%%%%%%%%%% FIGURE A4 %%%%%%%%%%%%%%%%%%%%%%%%%%%
%%%%%%%%%%%%%%%%%%%%%%%%%%%%%%%%%%%%%%%%%%%%%%%%%%%%

\subsection{Morphology and Kinematics of metallicity sub-components}
%%%%%%%%%%%%%%%%%% FIGURE A5 %%%%%%%%%%%%%%%%%%%%%%%%%%%
\begin{figure*}
\centering
\includegraphics[width=.75\textwidth]{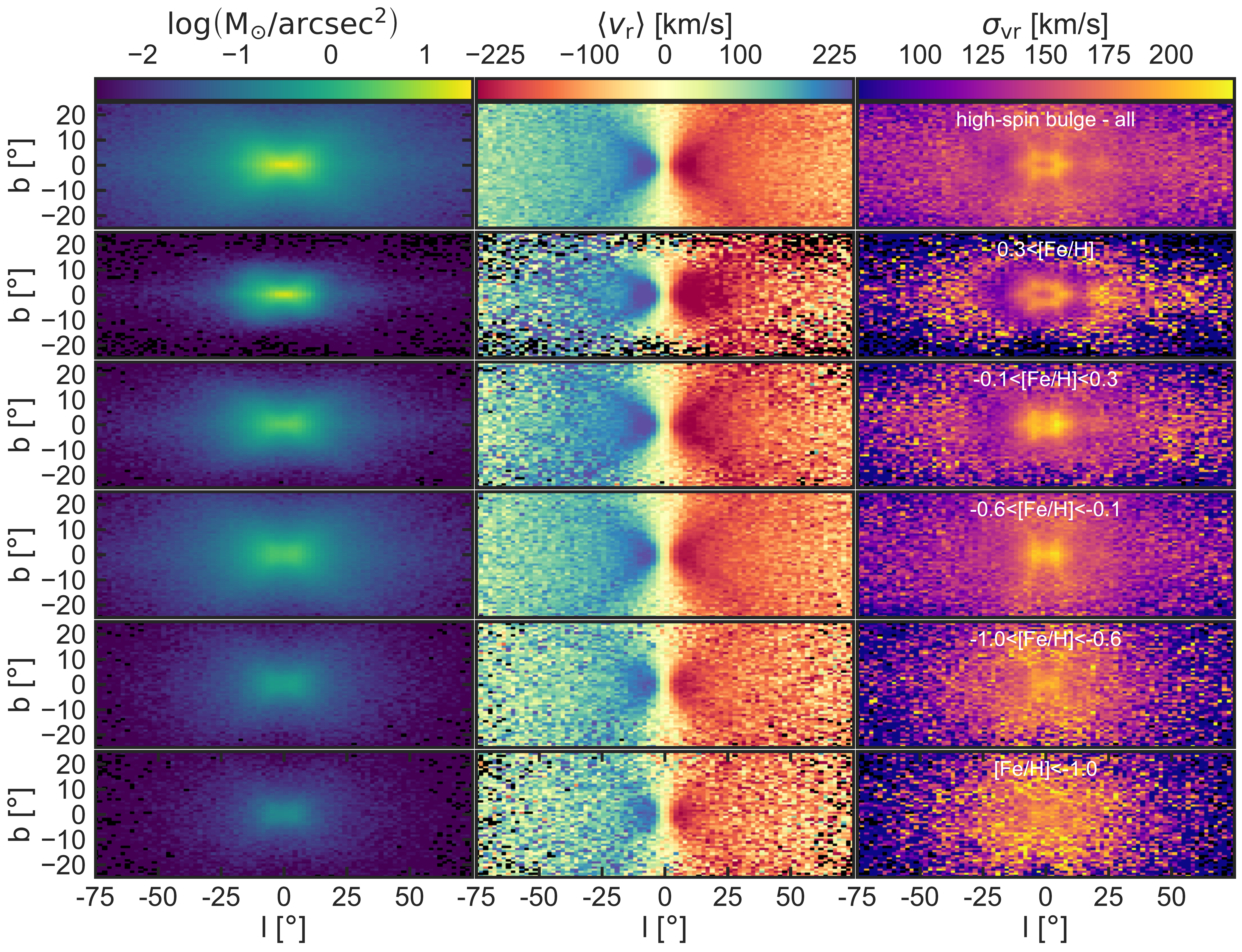}
\vspace{-.35cm}
\caption{Surface density (left column), mass-weighted rotation (middle column) and dispersion maps (right column) for different metallicity sub-components of the high-spin bulge stellar populations in $(l,b)$-projection for stars within $R=25$ kpc from the center of our simulated galaxy. The \emph{first row} shows the results for all stars of the high-spin bulge and then from top to bottom we show the components 1 to 5.
}
\label{fig:maps_decomp_pb}
\end{figure*}
%%%%%%%%%%%%%%%%%%%%%%%%%%%%%%%%%%%%%%%%%%%%%%%%%%%%

%%%%%%%%%%%%%%%%%% FIGURE A5 %%%%%%%%%%%%%%%%%%%%%%%%%%%
\begin{figure*}
\centering
\includegraphics[width=.75\textwidth]{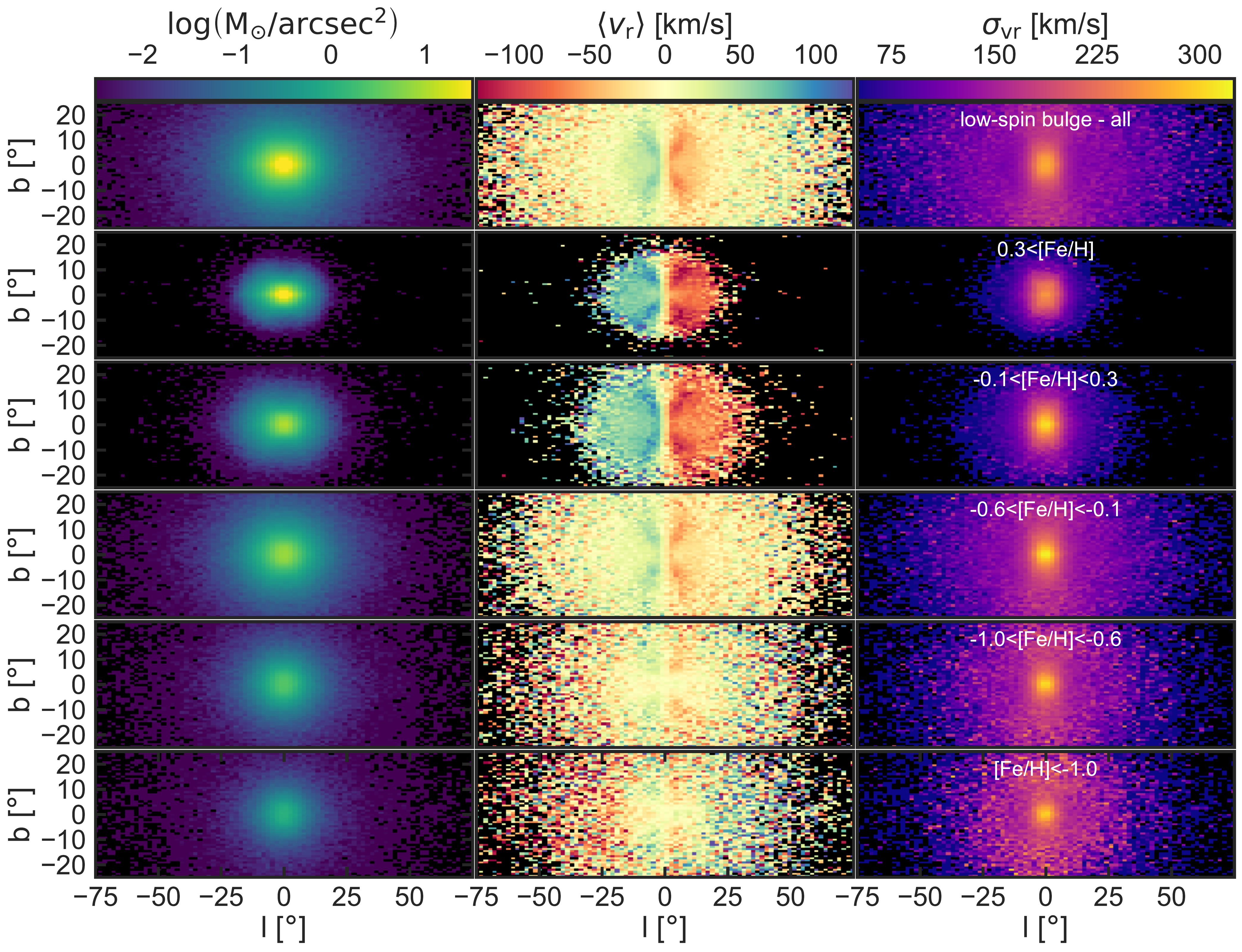}
\vspace{-.35cm}
\caption{Surface density (left column), mass-weighted rotation (middle column) and dispersion maps (right column) for different metallicity sub-components of the low-spin bulge stellar populations in $(l,b)$-projection for stars within $R=25$ kpc from the center of our simulated galaxy.  The \emph{first row} shows the results for all stars of the low-spin bulge and then from top to bottom we show the components 1 to 5.
}
\label{fig:maps_decomp_cb}
\end{figure*}
%%%%%%%%%%%%%%%%%%%%%%%%%%%%%%%%%%%%%%%%%%%%%%%%%%%%

%%%%%%%%%%%%%%%%%% FIGURE A6 %%%%%%%%%%%%%%%%%%%%%%%%%%%
\begin{figure*}
\centering
\includegraphics[width=.75\textwidth]{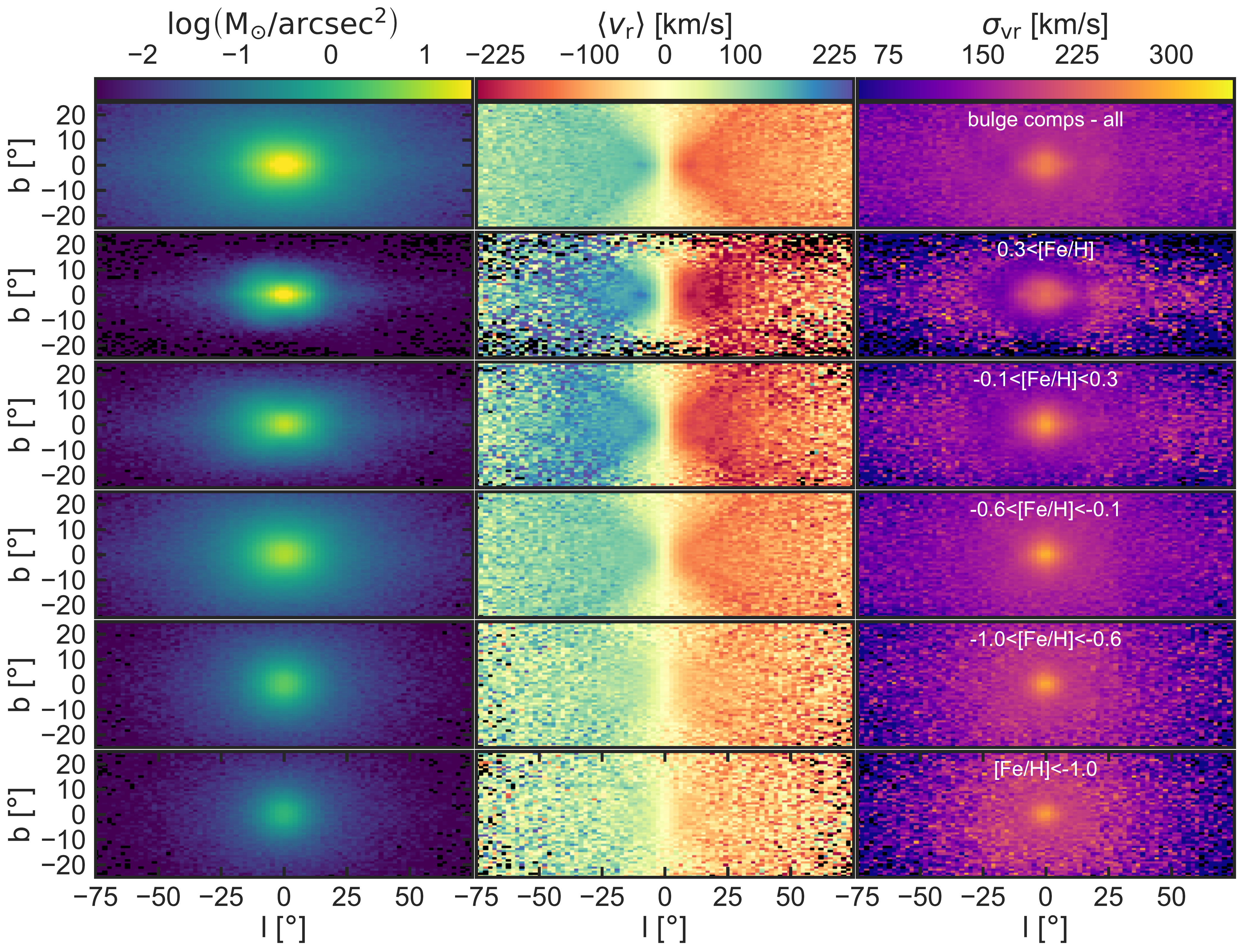}
\vspace{-.35cm}
\caption{Surface density (left column), mass-weighted rotation (middle column) and dispersion maps (right column) for different metallicity sub-components of the whole bulge stellar populations in $(l,b)$-projection for stars within $R=25$ kpc from the center of our simulated galaxy.  The \emph{first row} shows the results for all stars in the bulge and then from top to bottom we show the components 1 to 5.}
\label{fig:maps_decomp_both}
\end{figure*}
%%%%%%%%%%%%%%%%%%%%%%%%%%%%%%%%%%%%%%%%%%%%%%%%%%%%

\end{document}